\documentclass[10pt,journal,compsoc]{IEEEtran}
\ifCLASSOPTIONcompsoc
  \usepackage[nocompress]{cite}
\else
  \usepackage{cite}
\fi

\usepackage[pdftex]{graphicx}
\DeclareGraphicsExtensions{.pdf,.jpeg,.png}

\usepackage{amssymb}
\usepackage{array}
\ifCLASSOPTIONcompsoc
 \usepackage[caption=false,font=footnotesize,labelfont=sf,textfont=sf]{subfig}
\else
 \usepackage[caption=false,font=footnotesize]{subfig}
\fi

\usepackage{fixltx2e}
\usepackage{url}

\newcommand\MYhyperrefoptions{bookmarks=true,bookmarksnumbered=true,
pdfpagemode={UseOutlines},plainpages=false,pdfpagelabels=true,
colorlinks=true,linkcolor={black},citecolor={black},urlcolor={black},
pdftitle={A Survey of Software Foundations in Open Source},
pdfsubject={Manuscript},
pdfauthor={Javier L. C\'anovas Izquierdo, Jordi Cabot},
pdfkeywords={Open Source Software, Software Foundations}}

\ifCLASSINFOpdf
\usepackage[\MYhyperrefoptions,pdftex]{hyperref}
\else
\usepackage[\MYhyperrefoptions,breaklinks=true,dvips]{hyperref}
\usepackage{breakurl}
\fi

\usepackage{enumitem}
\usepackage{multirow}
\usepackage{booktabs}
\usepackage{framed}
\usepackage{threeparttable}
\usepackage{longtable}
\usepackage{lineno}
\usepackage{rotating}
\usepackage{tabularx}

\begin{document}
\title{A Survey of Software Foundations in \\ Open Source}
\author{Javier~Luis~C\'anovas~Izquierdo, Jordi~Cabot
\IEEEcompsocitemizethanks{\IEEEcompsocthanksitem J. L. C\'anovas Izquierdo and J. Cabot are with Universitat Oberta de Catalunya (UOC) -- Internet Interdisciplinary Institute (IN3).\protect\\
E-mail: jcanovasi@uoc.edu, jordi.cabot@icrea.cat
\IEEEcompsocthanksitem J. Cabot is with ICREA.}
\thanks{}}
\markboth{}%
{C\'anovas Izquierdo \MakeLowercase{\textit{et al.}}: A Survey of Software Foundations in Open Source}

\IEEEtitleabstractindextext{%
\begin{abstract}
A number of software foundations have been created as legal instruments to better articulate the structure, collaboration and financial model of Open Source Software (OSS) projects. 
Some examples are the Apache, Linux, or Mozilla foundations.
However, the mission and support provided by these foundations largely differ among them.
In this paper we perform a study on the role of foundations in OSS development. 
We analyze the nature, activities, role and governance of 101 software foundations and then go deeper on the 27 having as concrete goal the development and evolution of specific open source projects (and not just generic actions to promote the free software movement or similar).
Our results reveal the existence of a significant number of foundations with the sole purpose of promoting the free software movement and/or that limit themselves to core legal aspects but do not play any role in the day-to-day operations of the project (e.g., umbrella organizations for a large variety of projects). 
Therefore, while useful, foundations do not remove the need for specific projects to develop their own specific governance, contribution and development policies. 
A website to help projects to choose the foundation that best fits their needs is also available.
\end{abstract}

}

\maketitle
\IEEEdisplaynontitleabstractindextext

\IEEEpeerreviewmaketitle

\newlength{\mysep}
\setlength{\mysep}{0.35em}
\setlist[itemize]{leftmargin=0.3cm,labelindent=0cm}

\ifCLASSOPTIONcompsoc
\IEEEraisesectionheading{\section{Introduction}\label{sec:introduction}}
\else
\section{Introduction}
\label{sec:introduction}
\fi
Our society is undergoing a profound digital transformation with software playing a key role as the underlying digital infrastructure. 
Most of this software is (or heavily relies on) Open Source Software (OSS), thus contributing to this transformation and becoming a relevant part of the software industry and advancing the state of the art in research, education and government~(\cite{europe2015}).

Most OSS projects rely on active contributions of passionate developers to evolve~(\cite{Sen2012,Lee2009}). 
Therefore, long-term sustainability of an OSS project largely depends on its ability to retain developers and onboard new ones (i.e., newcomers), as well as to create a community of users who promote its adoption and use~(\cite{Bird2007,Zhou2015}).
As these projects grow, developers tend to organize and create communities to drive their development~(\cite{Allaho2013,OMahony2007,Crowston2007}) but many projects lack formal models, especially governance models~(\cite{IzquierdoC15}), to structure and manage the (potentially large) community around them. 
According to Laat's model~(\cite{Laat2007}) the structural evolution of OSS projects pass through three main phases, starting  from spontaneous coordination and ending into some kind of institutionalization.
Support to deal with all kinds of organizational decisions (including legal and economical aspects) is a huge concern for all projects at any stage.\looseness-1

In other domains, non-profit initiatives organize themselves around foundations (either public or private) that provide the legal and economical infrastructure for a community. 
Foundations can also define a number of internal regulations regarding, for instance, the activity, membership and decision-making process for the non-profit and non-governmental organizations. 
In the last years, we have also seen a number of foundations created around open source projects as a response to their maturation~(\cite{DBLP:conf/oss/RiehleB12}). 
Software foundations are non-profit organizations whose mission is to provide the needed grounds for open and collaborative software development. 
They also provide a legal framework for individual volunteers and enable the donation of resources for the public benefit~(\cite{DBLP:journals/computer/Riehle10}). 
However, there is a high variety of services and development strategies offered by software foundations.
For instance, the \emph{Apache Software Foundation} and the \emph{Linux Foundation} are two of the most well-known software foundations but they follow different strategies for the management of the projects they cover.
While the \emph{Apache Software Foundation} proposes a meritocratic system where different committees control and drive the development of several software projects (and the board oversees the whole process); the \emph{Linux Foundation} follows a flexible approach which serves as an umbrella for its projects, which can deploy specific development processes, and therefore the foundation itself concentrates more on the promotion of OSS benefits. 

In this paper we perform a survey of foundations behind OSS and their impact in the development of OSS projects. 
Our goal is to give a clear picture of the current state of the art of software foundations and to help developers make informed decisions when creating new ones or choosing an existing one to join.
To this aim, we build a dataset of 101 software foundations, which we analyze according to their nature, mission, openness and influence in the development practices taking place in the project itself.

The first version of this study was presented in a previous work (\cite{DBLP:conf/icse/IzquierdoC18}). 
This paper builds upon the previous work and extends it as follows: 
\begin{enumerate}[label=\Alph*]
\item The foundations dataset has been updated and includes 12 new foundations, thus resulting in a total of 101 foundations. 
\item The research method validation has been strengthened with a double coding schema to minimize internal validation threats.
\item We reorganized the research questions.
We now apply the first research questions to all software foundations (instead of filtering out some of them from the beginning), thus providing a more complete overview of the current state of such foundations. 
Also, we now have a specific question to classify foundations based on whether they actually support specific products or just focus on general evangelization of the free software movement (being the former the most useful ones for new projects looking for a foundation to join to).
\item We explored new subtopics in some of the research questions (e.g. the availability of a code of conduct at the foundation level).
\item We have developed a website summarizing the results of the study to increase its visibility and help other developers to choose the most suitable foundation for their projects. This website has also been useful to collect additional comments and feedback on the study that have been integrated in this extended version. 
\end{enumerate}

The paper is organized as follows.
Section \ref{sec:method} presents the methodology followed in our study, including the definition of the research questions and the construction of the dataset for our study.
Section \ref{sec:results} addresses the research questions and describes the results of our study.
Section \ref{sec:threats} describes the threats to validity.
Section \ref{sec:relatedWork} presents the related work.
Section \ref{sec:conclusion} finalizes the paper and presents the future work.

\section{Research Method}
\label{sec:method}
In this section we discuss how our study has been set up.
We first present our research questions (Section \ref{sec:method:rq}) and then describe how we built the dataset for our study (Section \ref{sec:method:dataset}). 

\subsection{Research Questions}
\label{sec:method:rq}
Our objective is to grasp a better understanding of the role played by software foundations in OSS development.
We identify the following main research questions:

\setlist[description]{leftmargin=0.2cm,labelindent=0cm}
\begin{description}
\item[RQ1] \textbf{What is the nature of software foundations?}
We perform a profiling analysis of software foundations in terms of their geographical scope, coverage and transparency (understood as the availability of explicit documents stating the foundation mission, activities and/or bylaws).

\item[RQ2] \textbf{What are the main activities of software foundations?}
We study the different ways software foundations help OSS (e.g., legal support, marketing or nurturing communities).
\item[RQ3] \textbf{How many software foundations actually support the development of a software product/s?}
We are especially interested in identifying those ones aiming to endorse and support specific OSS projects (e.g., instead of just being devoted to evangelization actions promoting the overall benefits of free software).
\end{description}


The previous research questions allow us to identify foundations that are international, independent and transparent, and that have the goal of supporting the development of software products. 
We believe these are the foundations that may have a larger impact on open source software and the ones that are especially useful for new OSS projects looking for a foundation to be part of. 
Therefore, we perform a further analysis of these foundations, in particular, by addressing the following research questions:

\begin{description}
\item[RQ4] \textbf{Do foundations play a direct role or have some influence in the way software development is actually conducted (i.e., governed)?}
There are many ways to support a software project (marketing, legal, infrastructure, etc.). 
As such, we analyze whether foundations limit themselves to provide a legal coverage for the projects or can be used by projects as a way to structure their own governance and contribution processes in detail (e.g., by giving templates or guidelines the projects can instantiate and even monitoring how well the projects adhere to them).

\item[RQ5] \textbf{How open are foundations developing software?}
We evaluate the openness of the foundations by examining their structure, how easy it is to become a member or how decisions are made. 
This can be an important criteria for projects looking for a suitable foundation (e.g., a project may favour more open foundations where project members could have an active participation at the foundation level vs. a more closed foundation where the project could benefit from the foundation support but not have any capacity to influence the foundation future direction).
\end{description}

\subsection{Dataset Construction}
\label{sec:method:dataset}

To answer the questions above, we built a dataset composed of a number of foundations.
To construct the dataset, we initially relied on the list of foundations available in \texttt{flossfoundations.org}\footnote{\texttt{http://flossfoundations.org}}, an online community created in 2005 by representatives from the \emph{Python Foundation}, \emph{Apache Software Foundation}, \emph{Perl Foundation} and \emph{Free Software Foundation}, 
with the aim of sharing their experiences and expertise in the field of free software and foundations.
We complemented this list by incorporating additional foundations from sources like \texttt{opensource.com}\footnote{\texttt{http://opensource.com}} and \texttt{oss-watch}\footnote{\texttt{http://oss-watch.ac.uk}}, as well as our own expertise.

For each foundation, we extracted 
(a) its URL; 
(b) the legal organization type (to verify that the foundation is a nonprofit organization); 
(c) the number of projects they cover (if any and publicly available); 
and (d) a brief description. When needed, additional information is later on retrieved to answer specific aspects of our research questions. 

At the end of the process, we constructed a dataset composed of 101 foundations. 
Table \ref{tab:foundations-rq1} (see first two columns) shows the list of foundations we collected reporting only aspects (a) and (c) due to space limitations.

\begin{figure}[t]
\centering
\includegraphics{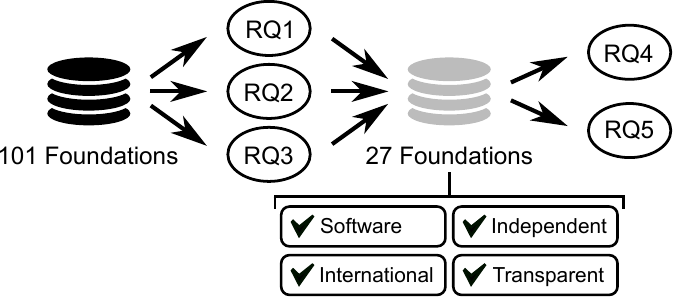}
\caption{Method followed to answer the research questions.}
\label{fig:process}
\end{figure}

According to the configuration of our method, RQ1, RQ2 and RQ3 receive the full set of foundations as input to study their nature and activities, respectively.
RQ4 and RQ5 take the international, independent and transparent foundations devoted to develop open source projects identified in RQ1, RQ2 and RQ3 to perform a deeper study. 
Figure~\ref{fig:process} shows the followed method and the size of the dataset involved in each research question.

\begin{table*}[p]
\centering
\begin{threeparttable}
\caption{Nature of Software Foundations.}
\label{tab:foundations-rq1}
\vspace{-2em}
\fontsize{5.65pt}{5.8pt}\selectfont
\begin{center}
\begin{tabularx}{0.7\textwidth}{l@{\hspace{0.5em}}X@{\hspace{0.5em}}r@{\hspace{0.65em}}c@{\hspace{0.5em}}c@{\hspace{0.5em}}c@{\hspace{0.5em}}} 
\toprule
\textsc{Name} & \textsc{Website URL} & \textsc{Size} & {\textsc{Sc}} & {\textsc{Co}} & {\textsc{Tr}}\\
\midrule
ADA Initiative                                      \dotfill & https://adainitiative.org/                                    \dotfill &     & $\blacksquare$ & $\blacksquare$ & $\blacksquare$ \\
Apache Software Foundation                          \dotfill & http://www.apache.org/foundation/                             \dotfill & 312 & $\blacksquare$ & $\blacksquare$ & $\blacksquare$ \\
Associacao SoftwareLivre.org                        \dotfill & http://associacao.softwarelivre.org/                          \dotfill &     & $\times$       & $\blacksquare$ & $\blacksquare$ \\
Benetech                                            \dotfill & http://www.benetech.org/                                      \dotfill &     & $\blacksquare$ & $\blacksquare$ & $\times$       \\
BioBricks Foundation                                \dotfill & https://biobricks.org/                                        \dotfill &     & $\blacksquare$ & $\blacksquare$ & $\times$       \\
Blender Foundation                                  \dotfill & http://www.blender.org/blenderorg/blender-foundation/         \dotfill & 1   & $\blacksquare$ & $\blacksquare$ & $\times$       \\
BSD Fund                                            \dotfill & http://bsdfund.org/                                           \dotfill & 4   & $\blacksquare$ & $\times$       & $\blacksquare$ \\
Cloud Foundry Foundation                            \dotfill & https://www.cloudfoundry.org/foundation/                      \dotfill & 1   & $\blacksquare$ & $\blacksquare$ & $\blacksquare$ \\
Creative Commons                                    \dotfill & http://creativecommons.org/                                   \dotfill &     & $\blacksquare$ & $\blacksquare$ & $\blacksquare$ \\
Digital Freedom Foundation                          \dotfill & https://www.digitalfreedomfoundation.org/                     \dotfill &     & $\blacksquare$ & $\times$       & $\times$       \\
Digital Freedom Foundation India                    \dotfill & http://dff.org.in/                                            \dotfill &     & $\times$       & $\blacksquare$ & $\times$       \\
Django Software Foundation                          \dotfill & http://www.djangoproject.com/foundation/                      \dotfill & 1   & $\blacksquare$ & $\blacksquare$ & $\blacksquare$ \\
Document Foundation                                 \dotfill & http://www.documentfoundation.org/                            \dotfill & 1   & $\blacksquare$ & $\blacksquare$ & $\blacksquare$ \\
Dojo Foundation                                     \dotfill & http://dojofoundation.org/                                    \dotfill & 1   & $\blacksquare$ & $\times$       & $\times$       \\
.NET Foundation                                     \dotfill & https://dotnetfoundation.org/                                 \dotfill & 556 & $\blacksquare$ & $\blacksquare$ & $\blacksquare$ \\
Eclipse Foundation                                  \dotfill & http://www.eclipse.org/org/foundation/                        \dotfill & 216 & $\blacksquare$ & $\blacksquare$ & $\blacksquare$ \\
El Centro de Software Libre                         \dotfill & http://www.csol.org/                                          \dotfill &     & $\times$       & $\blacksquare$ & $\times$       \\
Electronic Frontier Foundation                      \dotfill & http://www.eff.org/                                           \dotfill &     & $\blacksquare$ & $\blacksquare$ & $\times$       \\
Fintech Open Source Foundation                      \dotfill & https://www.finos.org/                                        \dotfill & 1   & $\blacksquare$ & $\blacksquare$ & $\blacksquare$ \\
Free Knowledge Institute                            \dotfill & http://freeknowledge.eu/                                      \dotfill &     & $\blacksquare$ & $\blacksquare$ & $\times$       \\
Free Software and Open Source Foundation for Africa \dotfill & http://www.fossfa.net/                                        \dotfill &     & $\times$       & $\blacksquare$ & $\blacksquare$ \\
Free Software Foundation                            \dotfill & http://www.fsf.org/                                           \dotfill &     & $\blacksquare$ & $\blacksquare$ & $\blacksquare$ \\
Free Software Foundation Europe                     \dotfill & https://fsfe.org                                              \dotfill &     & $\blacksquare$ & $\blacksquare$ & $\blacksquare$ \\
Free Software Foundation India                      \dotfill & http://fsf.org.in/                                            \dotfill &     & $\times$       & $\blacksquare$ & $\times$       \\
Free Software Foundation Latin America              \dotfill & http://www.fsfla.org/                                         \dotfill &     & $\times$       & $\blacksquare$ & $\blacksquare$ \\
FreeBSD Foundation                                  \dotfill & http://freebsdfoundation.org/                                 \dotfill & 1   & $\blacksquare$ & $\blacksquare$ & $\blacksquare$ \\
Fundación Vía Libre                                 \dotfill & http://www.vialibre.org.ar/                                   \dotfill &     & $\times$       & $\blacksquare$ & $\times$       \\
F\# Foundation                                      \dotfill & http://foundation.fsharp.org/                                 \dotfill & 1   & $\blacksquare$ & $\blacksquare$ & $\blacksquare$ \\
Gentoo Foundation                                   \dotfill & https://www.gentoo.org/inside-gentoo/foundation/              \dotfill & 1   & $\blacksquare$ & $\blacksquare$ & $\blacksquare$ \\
GNOME Foundation                                    \dotfill & http://foundation.gnome.org/                                  \dotfill & 1   & $\blacksquare$ & $\blacksquare$ & $\blacksquare$ \\
GraphQL Foundation                                  \dotfill & https://foundation.graphql.org/                               \dotfill & 1   & $\blacksquare$ & $\times$       & $\times$       \\
Kuali Foundation                                    \dotfill & https://kuali.org                                             \dotfill & 4   & $\blacksquare$ & $\blacksquare$ & $\blacksquare$ \\
Identity Commons                                    \dotfill & http://idcommons.net/                                         \dotfill &     & $\blacksquare$ & $\blacksquare$ & $\blacksquare$ \\
Internet Systems Consortium                         \dotfill & http://www.isc.org/                                           \dotfill & 9   & $\blacksquare$ & $\blacksquare$ & $\times$       \\
ITPUG (Italian PostgreSQL Users' Group)             \dotfill & https://www.itpug.org/                                        \dotfill & 1   & $\times$       & $\blacksquare$ & $\times$       \\
JS Foundation                                       \dotfill & https://js.foundation                                         \dotfill & 1   & $\blacksquare$ & $\times$       & $\blacksquare$ \\
JPUG (Japanese PostgreSQL Users' Group)             \dotfill & http://postgresql.jp                                          \dotfill & 1   & $\times$       & $\blacksquare$ & $\blacksquare$ \\
KDE e.V.                                            \dotfill & https://www.postgresql.jp/                                    \dotfill &     & $\blacksquare$ & $\blacksquare$ & $\blacksquare$ \\
Linux Expo of Southern California                   \dotfill & http://www.socallinuxexpo.org/                                \dotfill &     & $\blacksquare$ & $\blacksquare$ & $\times$       \\
Linux Foundation                                    \dotfill & http://linuxfoundation.org/                                   \dotfill & 89  & $\blacksquare$ & $\times$       & $\blacksquare$ \\
Linux Fund                                          \dotfill & http://www.linuxfund.org/                                     \dotfill & 4   & $\blacksquare$ & $\times$       & $\blacksquare$ \\
Linux International                                 \dotfill & http://www.li.org/                                            \dotfill &     & $\blacksquare$ & $\times$       & $\times$       \\
Linux Profesional Institute                         \dotfill & https://www.lpi.org/                                          \dotfill &     & $\blacksquare$ & $\blacksquare$ & $\times$       \\
LogiLogi Foundation                                 \dotfill & http://foundation.logilogi.org/                               \dotfill & 1   & $\blacksquare$ & $\blacksquare$ & $\times$       \\
Mambo Foundation  Inc.                              \dotfill & http://mambo-foundation.org/                                  \dotfill & 1   & $\blacksquare$ & $\blacksquare$ & $\times$       \\
Mozilla Foundation                                  \dotfill & http://www.mozilla.org/foundation/                            \dotfill & 9   & $\blacksquare$ & $\blacksquare$ & $\blacksquare$ \\
NetBSD Foundation                                   \dotfill & https://www.netbsd.org/foundation/                            \dotfill & 1   & $\blacksquare$ & $\blacksquare$ & $\blacksquare$ \\
NLnet Foundation                                    \dotfill & http://www.nlnet.nl/                                          \dotfill &     & $\blacksquare$ & $\blacksquare$ & $\blacksquare$ \\
NLnet Labs Foundation                               \dotfill & http://www.nlnetlabs.nl                                       \dotfill & 42  & $\blacksquare$ & $\blacksquare$ & $\blacksquare$ \\
NumFocus Inc.                                       \dotfill & https://www.numfocus.org/                                     \dotfill &     & $\blacksquare$ & $\blacksquare$ & $\blacksquare$ \\
One Laptop Per Child Association  Inc.              \dotfill & http://www.laptop.org/                                        \dotfill &     & $\blacksquare$ & $\blacksquare$ & $\times$       \\
Open Bioinformatics Foundation                      \dotfill & https://www.open-bio.org                                      \dotfill & 24  & $\blacksquare$ & $\times$       & $\blacksquare$ \\
Open Hardware Foundation                            \dotfill & http://www.openhardwarefoundation.org/                        \dotfill &     & $\blacksquare$ & $\blacksquare$ & $\times$       \\
Open Health Tools                                   \dotfill & http://wiki.p2pfoundation.net/Open\_Hardware\_Foundation      \dotfill & 36  & $\blacksquare$ & $\blacksquare$ & $\times$       \\
Open Media Now! Foundation                          \dotfill & http://www.openmedianow.org/                                  \dotfill &     & $\blacksquare$ & $\blacksquare$ & $\times$       \\
Open Source Applications Foundation                 \dotfill & http://www.osafoundation.org/                                 \dotfill & 1   & $\blacksquare$ & $\blacksquare$ & $\times$       \\
OSET Institute's                                    \dotfill & http://www.osdv.org/                                          \dotfill & 1   & $\blacksquare$ & $\blacksquare$ & $\times$       \\
Open Source For America                             \dotfill & http://www.osetfoundation.org                                 \dotfill &     & $\times$       & $\blacksquare$ & $\times$       \\
Open Source Geospatial Foundation                   \dotfill & http://www.osgeo.org/content/foundation/about.html            \dotfill & 32  & $\blacksquare$ & $\blacksquare$ & $\blacksquare$ \\
Open Source Initiative                              \dotfill & http://opensource.org/                                        \dotfill &     & $\blacksquare$ & $\blacksquare$ & $\blacksquare$ \\
Open Source Software Institute                      \dotfill & http://www.ossinstitute.org/                                  \dotfill &     & $\blacksquare$ & $\blacksquare$ & $\times$       \\
OpenBSD Foundation                                  \dotfill & http://openbsdfoundation.org/                                 \dotfill & 7   & $\blacksquare$ & $\blacksquare$ & $\blacksquare$ \\
Open Education Consortium                           \dotfill & http://www.oeconsortium.org/about-oec/                        \dotfill &     & $\blacksquare$ & $\blacksquare$ & $\blacksquare$ \\
OpenDoc Society                                     \dotfill & http://www.opendocsociety.org/                                \dotfill &     & $\blacksquare$ & $\blacksquare$ & $\blacksquare$ \\
OpenID Foundation                                   \dotfill & http://openid.net/foundation/                                 \dotfill &     & $\blacksquare$ & $\blacksquare$ & $\times$       \\
OpenSourceMatters                                   \dotfill & http://www.opensourcematters.org/                             \dotfill & 1   & $\blacksquare$ & $\blacksquare$ & $\blacksquare$ \\
OpenStack Foundation                                \dotfill & https://www.openstack.org/foundation/                         \dotfill & 6   & $\blacksquare$ & $\blacksquare$ & $\blacksquare$ \\
OpenStreetMap Foundation                            \dotfill & https://wiki.osmfoundation.org/wiki/Main\_Page                \dotfill & 1   & $\blacksquare$ & $\blacksquare$ & $\blacksquare$ \\
Oregon State University Open Source Lab Alliance    \dotfill & http://osuosl.org/                                            \dotfill & 160 & $\blacksquare$ & $\blacksquare$ & $\times$       \\
Parrot Foundation                                   \dotfill & http://www.parrot.org/foundation/                             \dotfill & 1   & $\blacksquare$ & $\blacksquare$ & $\blacksquare$ \\
Participatory Culture Foundation                    \dotfill & http://pculture.org/                                          \dotfill &     & $\blacksquare$ & $\blacksquare$ & $\times$       \\
Peer-Directed Projects Center (freenode)            \dotfill & https://en.wikipedia.org/wiki/Freenode                        \dotfill &     & $\blacksquare$ & $\blacksquare$ & $\times$       \\
Plone Foundation                                    \dotfill & http://plone.org/foundation                                   \dotfill & 1   & $\blacksquare$ & $\blacksquare$ & $\blacksquare$ \\
PostgreSQL Brasil                                   \dotfill & http://postgresql.org.br                                      \dotfill & 1   & $\times$       & $\blacksquare$ & $\times$       \\
PostgreSQL Europe                                   \dotfill & http://postgresql.eu                                          \dotfill & 1   & $\times$       & $\blacksquare$ & $\blacksquare$ \\
PostgreSQL.US                                       \dotfill & http://postgresql.us                                          \dotfill & 1   & $\times$       & $\blacksquare$ & $\blacksquare$ \\
PostgreSQLFr.org                                    \dotfill & http://asso.postgresql.fr                                     \dotfill & 1   & $\times$       & $\blacksquare$ & $\blacksquare$ \\
Public Software Fund                                \dotfill & http://www.pubsoft.org/                                       \dotfill & 43  & $\blacksquare$ & $\blacksquare$ & $\times$       \\
Python Software Foundation                          \dotfill & http://www.python.org/psf/                                    \dotfill & 1   & $\blacksquare$ & $\blacksquare$ & $\blacksquare$ \\
Sahana Foundation                                   \dotfill & https://sahanafoundation.org/                                 \dotfill & 1   & $\blacksquare$ & $\blacksquare$ & $\blacksquare$ \\
Shuttleworth Foundation                             \dotfill & http://www.shuttleworthfoundation.org/                        \dotfill &     & $\blacksquare$ & $\blacksquare$ & $\times$       \\
Software Freedom Conservancy                        \dotfill & http://sfconservancy.org/                                     \dotfill &     & $\blacksquare$ & $\blacksquare$ & $\blacksquare$ \\
Software Freedom Law Center                         \dotfill & http://www.softwarefreedom.org/                               \dotfill &     & $\blacksquare$ & $\blacksquare$ & $\times$       \\
Software in the Public Interest                     \dotfill & http://www.spi-inc.org/                                       \dotfill &     & $\blacksquare$ & $\blacksquare$ & $\blacksquare$ \\
Software Libre Argentina                            \dotfill & http://www.solarargentina.org/                                \dotfill &     & $\times$       & $\blacksquare$ & $\blacksquare$ \\
Software Libre Chile                                \dotfill & http://solarargentina.org/                                    \dotfill &     & $\times$       & $\blacksquare$ & $\times$       \\
Software Livre Brasil                               \dotfill & http://www.softwarelivre.org/                                 \dotfill &     & $\times$       & $\blacksquare$ & $\times$       \\
Subversion Corporation                              \dotfill & http://subversion.org/                                        \dotfill & 1   & $\blacksquare$ & $\times$       & $\blacksquare$ \\
TeX Users Group                                     \dotfill & http://tug.org/                                               \dotfill &     & $\blacksquare$ & $\blacksquare$ & $\blacksquare$ \\
The Open Planning Project                           \dotfill & http://theopenplanningproject.org/                            \dotfill & 1   & $\blacksquare$ & $\times$       & $\times$       \\
The Perl Foundation                                 \dotfill & http://perlfoundation.org/                                    \dotfill & 1   & $\blacksquare$ & $\blacksquare$ & $\blacksquare$ \\
The Software Conservancy                            \dotfill & http://www.tsc.org/                                           \dotfill &     & $\blacksquare$ & $\blacksquare$ & $\times$       \\
Twisted Software Foundation                         \dotfill & http://twistedmatrix.com/trac/wiki/TwistedSoftwareFoundation  \dotfill & 1   & $\blacksquare$ & $\times$       & $\times$       \\
TYPO3 Association                                   \dotfill & http://association.typo3.org/                                 \dotfill & 1   & $\blacksquare$ & $\blacksquare$ & $\times$       \\
Wikimedia Foundation                                \dotfill & http://www.wikimediafoundation.org                            \dotfill & 1   & $\blacksquare$ & $\blacksquare$ & $\blacksquare$ \\
Wikiotics Foundation                                \dotfill & https://wikiotics.org/en/Wikiotics\_Foundation                \dotfill & 1   & $\blacksquare$ & $\blacksquare$ & $\times$       \\
Wordpress Foundation                                \dotfill & http://wordpressfoundation.org                                \dotfill & 1   & $\blacksquare$ & $\blacksquare$ & $\times$       \\
X.Org Foundation LLC                                \dotfill & http://www.x.org/wiki/XorgFoundation                          \dotfill & 1   & $\blacksquare$ & $\blacksquare$ & $\blacksquare$ \\
Xiph.org                                            \dotfill & http://xiph.org/                                              \dotfill & 23  & $\blacksquare$ & $\blacksquare$ & $\times$       \\
XMPP Standards Foundation                           \dotfill & https://xmpp.org/about/xmpp-standards-foundation.html         \dotfill &     & $\blacksquare$ & $\blacksquare$ & $\blacksquare$ \\
Zope Foundation                                     \dotfill & http://foundation.zope.org/                                   \dotfill & 1   & $\blacksquare$ & $\blacksquare$ & $\times$       \\
\bottomrule
\end{tabularx}
\begin{tablenotes}
\item Legend: \textsc{Sc} = Scope. \textsc{Co} = Coverage. \textsc{Tr} = Transparency.
\item \textsc{Size} refers to the number of projects covered.
\end{tablenotes}
\end{center}
\end{threeparttable}
\end{table*}

\section{Results}
\label{sec:results}
In the following we address our research questions based on the analysis of the foundations considered in our dataset.


\subsection{RQ1: Nature of Software Foundations}
To address this research question, we study each foundation and create a profile about their nature, according to three orthogonal dimensions, namely: geographical scope, coverage and transparency.

\vspace{\mysep}
\noindent\textbf{Geographical Scope}. 
This dimension studies the distribution of the foundations from a geographical point of view (i.e., whether the foundation has an international or local character). 
We detected 17 software foundations that were mainly focused on the development of local OSS communities. 
For instance, \emph{ITPUG} or \emph{JPUG} are foundations whose activities focus on PostgreSQL projects developed in Italy and Japan, respectively.
We are mainly interested in foundations having an international distribution, as local ones are usually smaller and may adopt some behavior specific to the country and legislation where they are located.
Table~\ref{tab:foundations-rq1}, column \textsc{Sc}, marks with a black square the foundations with an international distribution (and a cross for the rest). 

\vspace{\mysep}
\noindent\textbf{Coverage}. 
Foundations can either serve a specific project, a set of projects or provide an umbrella for a number of smaller foundations that use it to simplify its own creation, management and legal processes.
We found 12 foundations that mainly act as umbrella for others, like the \emph{Linux Foundation}, where inner organizations generally work in a stand-alone way (e.g., \emph{OpenAPI initiative} or \emph{Kubernetes} within the \emph{Linux Foundation}); or as subsidiaries, like \emph{Subversion Corporation}, which is part of the \emph{Apache Software Foundation}.
We are interested in independent foundations (i.e., neither umbrella nor subsidiary foundations) as individual analysis subjects for the study. 
Table~\ref{tab:foundations-rq1}, column \textsc{Co}, signals our independent foundations with a black square.  

\vspace{\mysep}
\noindent\textbf{Transparency}. 
Foundations may aim to help OSS projects on several dimensions. These goals should be clearly stated as part of the foundations' documentation and mentioned either in the official foundation resources or in their bylaws.
Surprisingly, a significant number of foundations are not so transparent and did not provide any explicit information on this regard. 
According to our evaluation, 45 foundations did not include any reference to one or more of the following elements: mission, organization, projects or bylaws, being the last one the most difficult to find. 
Table~\ref{tab:foundations-rq1}, column \textsc{Tr}, shows a black square for those foundations which were transparent enough to provide the previous information.
For those, we collected the available information and performed a term frequency study.
Apart from the high frequency of terms directly related to their main tasks like \emph{software}, \emph{development}, \emph{open} or \emph{free}; the term \emph{community} is also included in the mission descriptions, thus revealing the importance of this dimension.
Other relevant terms with high frequency refer to the core promotion and support of OSS (e.g., \emph{promote}, \emph{support}, \emph{defend} or \emph{infrastructure}).

\begin{framed}
\noindent 83\% of the surveyed foundations have an international vocation and 88\% are independent single software foundations. 
For the 55\% of the software foundations with an explicit mission description, the community and defense of OSS are key concepts together with the development support.
\end{framed}


\subsection{RQ2: Main Activities of Software Foundations}

To address this research question we classified the foundations applying a grounded theory approach.
First, we analyzed the resources of each foundation (i.e., website, forums, official documents, etc.)\footnote{Note that we could perform this analysis for foundations not fully transparent (i.e., RQ1) even though in these cases the results were more partial as they were reconstructed from separate scattered pieces of information instead of having a single source of truth to rely on.}.
Second, we performed open coding on these results with the purpose of assigning them to specific topics.
Finally, we grouped the conceptually similar topics, resulting into eight main categories.

The identified categories are:
(1) OSS promotion, for those foundations aimed at promoting OSS and its ideas (i.e., evangelization);
(2) event organization, for those foundations which help to organize events in the context of OSS (e.g., developer meetups, project development meetings, etc.);
(3) project sponsoring, for those foundations supporting (mainly economically) OSS initiatives; 
(4) training, for those foundations offering courses about OSS practices ranging from development methods (e.g., how to deploy a pull request development process) to product-specific seminars (e.g., how to install Linux or \TeX);
(5) legal support, for those foundations which assist developers on legal issues of OSS projects (e.g., licensing or patents);
(6) government involvement, for those foundations which are involved in involving governmental entities into OSS (e.g., participating in the creation of regulations to promote OSS);
(7) standard leaders, for those foundations which help on defining standards or specifications; and
(8) community coordination, for those foundations responsible of nurturing and providing the means to manage communities.

\begin{table*}[p]
\caption{Activities of software foundations.}
\label{tab:foundations-rq2}
\vspace{-2em}
\fontsize{5.30pt}{5.8pt}\selectfont
\begin{center}
\begin{tabularx}{0.86\textwidth}{l@{\hspace{0.5em}}c@{\hspace{0.5em}}c@{\hspace{0.5em}}c@{\hspace{0.5em}}c@{\hspace{0.5em}}c@{\hspace{0.5em}}c@{\hspace{0.5em}}c@{\hspace{0.5em}}c@{\hspace{0.5em}}c@{\hspace{0.5em}}} 
\toprule
\multirow{2}{*}{\textsc{Name}} 
 & {\textsc{OSS}}       & {\textsc{Event}}        & {\textsc{Project}}    & \multirow{2}{*}{\textsc{Training}} & {\textsc{Legal}}   & {\textsc{Government}}  & {\textsc{Standard}} & {\textsc{Community}}    & {\textsc{Software}} \\
 & {\textsc{Promotion}} & {\textsc{Organization}} & {\textsc{Sponsoring}} &                                    & {\textsc{Support}} & {\textsc{Involvement}} & {\textsc{Leaders}}  & {\textsc{Coordination}} & {\textsc{Development}} \\
\midrule
ADA Initiative                                      \dotfill & ---            & $\blacksquare$ & ---            & $\blacksquare$ & $\blacksquare$ & ---            & ---            & $\blacksquare$ & ---            \\
Apache Software Foundation                          \dotfill & ---            & ---            & $\blacksquare$ & ---            & $\blacksquare$ & ---            & ---            & $\blacksquare$ & $\blacksquare$ \\
Associacao SoftwareLivre.org                        \dotfill & $\blacksquare$ & $\blacksquare$ & ---            & $\blacksquare$ & ---            & $\blacksquare$ & ---            & ---            & ---            \\
Benetech                                            \dotfill & ---            & ---            & $\blacksquare$ & ---            & ---            & ---            & ---            & $\blacksquare$ & ---            \\
BioBricks Foundation                                \dotfill & ---            & $\blacksquare$ & $\blacksquare$ & ---            & ---            & ---            & ---            & ---            & ---            \\
Blender Foundation                                  \dotfill & ---            & ---            & $\blacksquare$ & ---            & ---            & ---            & ---            & $\blacksquare$ & $\blacksquare$ \\
BSD Fund                                            \dotfill & ---            & $\blacksquare$ & $\blacksquare$ & ---            & ---            & ---            & ---            & $\blacksquare$ & $\blacksquare$ \\
Cloud Foundry Foundation                            \dotfill & ---            & $\blacksquare$ & $\blacksquare$ & $\blacksquare$ & ---            & ---            & ---            & $\blacksquare$ & $\blacksquare$ \\
Creative Commons                                    \dotfill & ---            & ---            & ---            & ---            & $\blacksquare$ & $\blacksquare$ & $\blacksquare$ & ---            & ---            \\
Digital Freedom Foundation                          \dotfill & $\blacksquare$ & ---            & ---            & $\blacksquare$ & $\blacksquare$ & ---            & ---            & ---            & ---            \\
Digital Freedom Foundation India                    \dotfill & $\blacksquare$ & $\blacksquare$ & $\blacksquare$ & $\blacksquare$ & ---            & ---            & ---            & ---            & ---            \\
Django Software Foundation                          \dotfill & ---            & ---            & $\blacksquare$ & ---            & $\blacksquare$ & ---            & ---            & $\blacksquare$ & $\blacksquare$ \\
Document Foundation                                 \dotfill & ---            & ---            & $\blacksquare$ & $\blacksquare$ & ---            & ---            & ---            & $\blacksquare$ & $\blacksquare$ \\
Dojo Foundation                                     \dotfill & ---            & $\blacksquare$ & $\blacksquare$ & $\blacksquare$ & ---            & ---            & ---            & $\blacksquare$ & $\blacksquare$ \\
.NET Foundation                                     \dotfill & $\blacksquare$ & $\blacksquare$ & $\blacksquare$ & ---            & $\blacksquare$ & ---            & ---            & $\blacksquare$ & $\blacksquare$ \\
Eclipse Foundation                                  \dotfill & ---            & $\blacksquare$ & $\blacksquare$ & ---            & ---            & ---            & ---            & $\blacksquare$ & $\blacksquare$ \\
El Centro de Software Libre                         \dotfill & $\blacksquare$ & $\blacksquare$ & ---            & ---            & ---            & $\blacksquare$ & ---            & $\blacksquare$ & ---            \\
Electronic Frontier Foundation                      \dotfill & ---            & ---            & ---            & ---            & $\blacksquare$ & $\blacksquare$ & ---            & ---            & ---            \\
Fintech Open Source Foundation                      \dotfill & ---            & $\blacksquare$ & $\blacksquare$ & ---            & ---            & ---            & ---            & $\blacksquare$ & $\blacksquare$ \\
Free Knowledge Institute                            \dotfill & $\blacksquare$ & ---            & ---            & ---            & ---            & ---            & ---            & $\blacksquare$ & ---            \\
Free Software and Open Source Foundation for Africa \dotfill & $\blacksquare$ & ---            & ---            & ---            & ---            & $\blacksquare$ & $\blacksquare$ & $\blacksquare$ & ---            \\
Free Software Foundation                            \dotfill & $\blacksquare$ & $\blacksquare$ & $\blacksquare$ & ---            & $\blacksquare$ & ---            & $\blacksquare$ & $\blacksquare$ & ---            \\
Free Software Foundation Europe                     \dotfill & $\blacksquare$ & $\blacksquare$ & $\blacksquare$ & $\blacksquare$ & $\blacksquare$ & $\blacksquare$ & $\blacksquare$ & $\blacksquare$ & ---            \\
Free Software Foundation India                      \dotfill & $\blacksquare$ & $\blacksquare$ & ---            & ---            & ---            & ---            & ---            & $\blacksquare$ & ---            \\
Free Software Foundation Latin America              \dotfill & $\blacksquare$ & ---            & ---            & $\blacksquare$ & $\blacksquare$ & $\blacksquare$ & ---            & $\blacksquare$ & ---            \\
FreeBSD Foundation                                  \dotfill & ---            & $\blacksquare$ & $\blacksquare$ & $\blacksquare$ & $\blacksquare$ & ---            & ---            & $\blacksquare$ & $\blacksquare$ \\
Fundación Vía Libre                                 \dotfill & $\blacksquare$ & $\blacksquare$ & ---            & ---            & ---            & ---            & ---            & ---            & ---            \\
F\# Foundation                                      \dotfill & ---            & ---            & $\blacksquare$ & ---            & ---            & ---            & ---            & $\blacksquare$ & $\blacksquare$ \\
Gentoo Foundation                                   \dotfill & ---            & $\blacksquare$ & $\blacksquare$ & $\blacksquare$ & ---            & ---            & ---            & $\blacksquare$ & $\blacksquare$ \\
GNOME Foundation                                    \dotfill & ---            & ---            & $\blacksquare$ & ---            & ---            & ---            & ---            & $\blacksquare$ & $\blacksquare$ \\
GraphQL Foundation                                  \dotfill & ---            & ---            & $\blacksquare$ & ---            & ---            & ---            & ---            & $\blacksquare$ & $\blacksquare$ \\
Kuali Foundation                                    \dotfill & ---            & $\blacksquare$ & $\blacksquare$ & ---            & ---            & ---            & ---            & $\blacksquare$ & $\blacksquare$ \\
Identity Commons                                    \dotfill & ---            & $\blacksquare$ & ---            & ---            & ---            & ---            & $\blacksquare$ & $\blacksquare$ & ---            \\
Internet Systems Consortium                         \dotfill & ---            & ---            & $\blacksquare$ & ---            & ---            & ---            & $\blacksquare$ & $\blacksquare$ & $\blacksquare$ \\
ITPUG (Italian PostgreSQL Users' Group)             \dotfill & ---            & $\blacksquare$ & ---            & $\blacksquare$ & ---            & ---            & ---            & $\blacksquare$ & $\blacksquare$ \\
JS Foundation                                       \dotfill & ---            & $\blacksquare$ & $\blacksquare$ & $\blacksquare$ & ---            & ---            & ---            & $\blacksquare$ & $\blacksquare$ \\
JPUG (Japanese PostgreSQL Users' Group)             \dotfill & ---            & $\blacksquare$ & ---            & $\blacksquare$ & ---            & ---            & ---            & $\blacksquare$ & $\blacksquare$ \\
KDE e.V.                                            \dotfill & ---            & $\blacksquare$ & $\blacksquare$ & $\blacksquare$ & ---            & ---            & ---            & $\blacksquare$ & ---            \\
Linux Expo of Southern California                   \dotfill & $\blacksquare$ & $\blacksquare$ & ---            & ---            & ---            & ---            & ---            & $\blacksquare$ & ---            \\
Linux Foundation                                    \dotfill & $\blacksquare$ & $\blacksquare$ & $\blacksquare$ & $\blacksquare$ & ---            & ---            & ---            & $\blacksquare$ & $\blacksquare$ \\
Linux Fund                                          \dotfill & ---            & ---            & $\blacksquare$ & ---            & ---            & ---            & ---            & $\blacksquare$ & $\blacksquare$ \\
Linux International                                 \dotfill & $\blacksquare$ & ---            & $\blacksquare$ & ---            & ---            & ---            & ---            & $\blacksquare$ & $\blacksquare$ \\
Linux Profesional Institute                         \dotfill & ---            & ---            & ---            & $\blacksquare$ & ---            & ---            & ---            & ---            & ---            \\
LogiLogi Foundation                                 \dotfill & $\blacksquare$ & ---            & ---            & ---            & ---            & ---            & ---            & $\blacksquare$ & ---            \\
Mambo Foundation  Inc.                              \dotfill & ---            & ---            & $\blacksquare$ & ---            & $\blacksquare$ & ---            & ---            & $\blacksquare$ & $\blacksquare$ \\
Mozilla Foundation                                  \dotfill & $\blacksquare$ & $\blacksquare$ & $\blacksquare$ & $\blacksquare$ & ---            & ---            & ---            & $\blacksquare$ & $\blacksquare$ \\
NetBSD Foundation                                   \dotfill & ---            & ---            & $\blacksquare$ & ---            & ---            & ---            & ---            & $\blacksquare$ & $\blacksquare$ \\
NLnet Foundation                                    \dotfill & ---            & $\blacksquare$ & $\blacksquare$ & ---            & ---            & ---            & ---            & ---            & ---            \\
NLnet Labs Foundation                               \dotfill & ---            & ---            & $\blacksquare$ & ---            & ---            & ---            & ---            & $\blacksquare$ & $\blacksquare$ \\
NumFocus Inc.                                       \dotfill & $\blacksquare$ & ---            & ---            & $\blacksquare$ & $\blacksquare$ & ---            & ---            & ---            & ---            \\
One Laptop Per Child Association  Inc.              \dotfill & $\blacksquare$ & ---            & ---            & $\blacksquare$ & ---            & ---            & ---            & ---            & ---            \\
Open Bioinformatics Foundation                      \dotfill & ---            & $\blacksquare$ & $\blacksquare$ & $\blacksquare$ & ---            & ---            & ---            & $\blacksquare$ & $\blacksquare$ \\
Open Hardware Foundation                            \dotfill & $\blacksquare$ & ---            & ---            & ---            & ---            & ---            & ---            & ---            & ---            \\
Open Health Tools                                   \dotfill & ---            & ---            & $\blacksquare$ & ---            & ---            & ---            & ---            & ---            & $\blacksquare$ \\
Open Media Now! Foundation                          \dotfill & $\blacksquare$ & ---            & ---            & ---            & ---            & ---            & ---            & ---            & ---            \\
Open Source Applications Foundation                 \dotfill & $\blacksquare$ & ---            & $\blacksquare$ & ---            & ---            & ---            & ---            & $\blacksquare$ & $\blacksquare$ \\
Open Source Digital Voting Foundation               \dotfill & ---            & ---            & $\blacksquare$ & ---            & ---            & ---            & ---            & ---            & $\blacksquare$ \\
Open Source For America                             \dotfill & $\blacksquare$ & ---            & ---            & ---            & ---            & $\blacksquare$ & ---            & $\blacksquare$ & ---            \\
Open Source Geospatial Foundation                   \dotfill & $\blacksquare$ & ---            & $\blacksquare$ & ---            & ---            & ---            & ---            & $\blacksquare$ & $\blacksquare$ \\
Open Source Initiative                              \dotfill & $\blacksquare$ & ---            & ---            & ---            & ---            & ---            & $\blacksquare$ & ---            & ---            \\
Open Source Software Institute                      \dotfill & $\blacksquare$ & ---            & $\blacksquare$ & ---            & ---            & ---            & ---            & ---            & ---            \\
OpenBSD Foundation                                  \dotfill & ---            & $\blacksquare$ & $\blacksquare$ & ---            & ---            & ---            & ---            & $\blacksquare$ & $\blacksquare$ \\
Open Education Consortium                           \dotfill & ---            & $\blacksquare$ & ---            & $\blacksquare$ & ---            & ---            & ---            & ---            & ---            \\
OpenDoc Society                                     \dotfill & $\blacksquare$ & $\blacksquare$ & ---            & $\blacksquare$ & ---            & ---            & ---            & $\blacksquare$ & ---            \\
OpenID Foundation                                   \dotfill & ---            & ---            & ---            & ---            & $\blacksquare$ & ---            & $\blacksquare$ & $\blacksquare$ & ---            \\
OpenSourceMatters                                   \dotfill & ---            & $\blacksquare$ & $\blacksquare$ & ---            & ---            & ---            & ---            & $\blacksquare$ & $\blacksquare$ \\
OpenStack Foundation                                \dotfill & ---            & $\blacksquare$ & $\blacksquare$ & $\blacksquare$ & ---            & ---            & ---            & $\blacksquare$ & $\blacksquare$ \\
OpenStreetMap Foundation                            \dotfill & ---            & ---            & $\blacksquare$ & ---            & $\blacksquare$ & ---            & ---            & $\blacksquare$ & $\blacksquare$ \\
Oregon State University Open Source Lab Alliance    \dotfill & $\blacksquare$ & $\blacksquare$ & $\blacksquare$ & ---            & ---            & ---            & ---            & ---            & $\blacksquare$ \\
Parrot Foundation                                   \dotfill & $\blacksquare$ & ---            & $\blacksquare$ & ---            & ---            & ---            & ---            & $\blacksquare$ & $\blacksquare$ \\
Participatory Culture Foundation                    \dotfill & ---            & ---            & $\blacksquare$ & ---            & ---            & ---            & ---            & $\blacksquare$ & ---            \\
Peer-Directed Projects Center (freenode)            \dotfill & ---            & ---            & ---            & ---            & ---            & ---            & ---            & $\blacksquare$ & ---            \\
Plone Foundation                                    \dotfill & ---            & $\blacksquare$ & $\blacksquare$ & $\blacksquare$ & ---            & ---            & ---            & $\blacksquare$ & $\blacksquare$ \\
PostgreSQL Brasil                                   \dotfill & ---            & $\blacksquare$ & $\blacksquare$ & ---            & ---            & ---            & ---            & $\blacksquare$ & $\blacksquare$ \\
PostgreSQL Europe                                   \dotfill & ---            & $\blacksquare$ & $\blacksquare$ & ---            & ---            & ---            & ---            & $\blacksquare$ & $\blacksquare$ \\
PostgreSQL.US                                       \dotfill & ---            & $\blacksquare$ & $\blacksquare$ & ---            & ---            & ---            & ---            & $\blacksquare$ & $\blacksquare$ \\
PostgreSQLFr.org                                    \dotfill & ---            & $\blacksquare$ & $\blacksquare$ & ---            & ---            & ---            & ---            & $\blacksquare$ & $\blacksquare$ \\
Public Software Fund                                \dotfill & ---            & ---            & $\blacksquare$ & ---            & ---            & ---            & ---            & ---            & $\blacksquare$ \\
Python Software Foundation                          \dotfill & ---            & $\blacksquare$ & $\blacksquare$ & ---            & ---            & ---            & ---            & $\blacksquare$ & $\blacksquare$ \\
Sahana Foundation                                   \dotfill & $\blacksquare$ & ---            & $\blacksquare$ & ---            & ---            & ---            & ---            & $\blacksquare$ & $\blacksquare$ \\
Shuttleworth Foundation                             \dotfill & $\blacksquare$ & ---            & $\blacksquare$ & ---            & ---            & ---            & ---            & ---            & ---            \\
Software Freedom Conservancy                        \dotfill & $\blacksquare$ & ---            & $\blacksquare$ & ---            & $\blacksquare$ & ---            & ---            & ---            & ---            \\
Software Freedom Law Center                         \dotfill & ---            & ---            & ---            & $\blacksquare$ & $\blacksquare$ & ---            & ---            & ---            & ---            \\
Software in the Public Interest                     \dotfill & $\blacksquare$ & ---            & $\blacksquare$ & ---            & ---            & ---            & ---            & ---            & ---            \\
Software Libre Argentina                            \dotfill & ---            & $\blacksquare$ & ---            & $\blacksquare$ & ---            & ---            & ---            & ---            & ---            \\
Software Libre Chile                                \dotfill & ---            & $\blacksquare$ & ---            & $\blacksquare$ & ---            & ---            & ---            & ---            & ---            \\
Software Livre Brasil                               \dotfill & ---            & $\blacksquare$ & ---            & $\blacksquare$ & ---            & ---            & ---            & ---            & ---            \\
Subversion Corporation                              \dotfill & ---            & ---            & $\blacksquare$ & ---            & ---            & ---            & ---            & $\blacksquare$ & $\blacksquare$ \\
TeX Users Group                                     \dotfill & ---            & ---            & ---            & $\blacksquare$ & ---            & ---            & ---            & $\blacksquare$ & ---            \\
The Open Planning Project                           \dotfill & ---            & ---            & ---            & ---            & ---            & ---            & ---            & ---            & $\blacksquare$ \\
The Perl Foundation                                 \dotfill & ---            & $\blacksquare$ & $\blacksquare$ & ---            & ---            & ---            & ---            & $\blacksquare$ & $\blacksquare$ \\
The Software Conservancy                            \dotfill & ---            & $\blacksquare$ & ---            & ---            & ---            & ---            & ---            & ---            & ---            \\
Twisted Software Foundation                         \dotfill & ---            & ---            & $\blacksquare$ & ---            & ---            & ---            & ---            & $\blacksquare$ & $\blacksquare$ \\
TYPO3 Association                                   \dotfill & ---            & ---            & $\blacksquare$ & ---            & ---            & ---            & ---            & $\blacksquare$ & $\blacksquare$ \\
Wikimedia Foundation                                \dotfill & $\blacksquare$ & $\blacksquare$ & $\blacksquare$ & ---            & $\blacksquare$ & ---            & ---            & $\blacksquare$ & $\blacksquare$ \\
Wikiotics Foundation                                \dotfill & ---            & ---            & $\blacksquare$ & ---            & ---            & ---            & ---            & $\blacksquare$ & $\blacksquare$ \\
Wordpress Foundation                                \dotfill & ---            & $\blacksquare$ & $\blacksquare$ & $\blacksquare$ & $\blacksquare$ & ---            & ---            & $\blacksquare$ & $\blacksquare$ \\
X.Org Foundation LLC                                \dotfill & ---            & ---            & $\blacksquare$ & ---            & ---            & ---            & ---            & $\blacksquare$ & $\blacksquare$ \\
Xiph.org                                            \dotfill & $\blacksquare$ & ---            & $\blacksquare$ & ---            & ---            & ---            & ---            & $\blacksquare$ & $\blacksquare$ \\
XMPP Standards Foundation                           \dotfill & ---            & ---            & ---            & ---            & ---            & ---            & $\blacksquare$ & ---            & ---            \\
Zope Foundation                                     \dotfill & ---            & ---            & $\blacksquare$ & ---            & ---            & ---            & ---            & $\blacksquare$ & $\blacksquare$ \\
\bottomrule& 
\end{tabularx}
\end{center}
\end{table*}

\begin{figure}[t]
\centering
\includegraphics{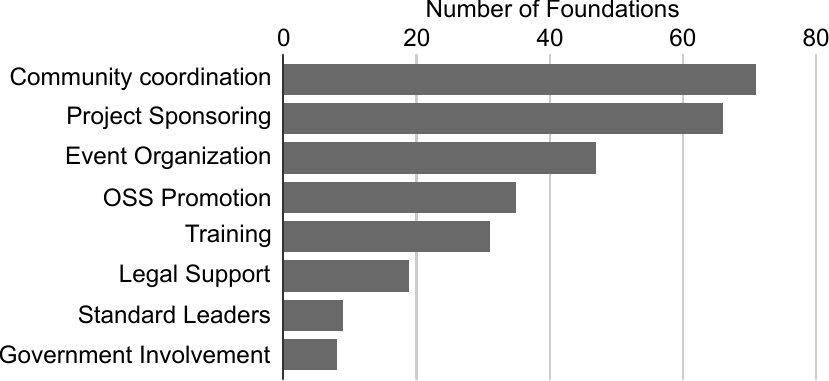}
\caption{Main activities distribution.}
\label{fig:topicPieChart}
\end{figure}

Figure~\ref{fig:topicPieChart} shows the distribution of these categories for the foundations of our study, while Table~\ref{tab:foundations-rq2} (first 9 columns) shows the results for each foundation of our dataset.
As can be seen, analyzed foundations have specific support for the promotion of OSS as a whole, event organization and taking care of the community, thus facilitating the marketing and communication of the benefits of the open and free software movement. 

\begin{framed}
\noindent Surveyed foundations devote specific support for the promotion of open and free software movement and the management of user/developers communities.
\end{framed}


\subsection{RQ3: Foundations Focused on Creating Software}
We believe that the results of the main activities of software foundations reported in RQ2 are crucial for the sustainability of open source communities, however, we want to focus on those foundations specifically aimed to support the development of OSS projects. Thus to address this research question we analyzed each foundation to check if they are directly involved in endorsing the development of a software product/s.

We present the results in Table~\ref{tab:foundations-rq2} (see last column) in order to facilitate comparison with the results of RQ2.
As can be seen, most analyzed foundations have as main activity the support of specific OSS projects. 
In comparison to results of RQ2, those that do not have this goal are mainly focused on the promotion of promotion of open and free software movement, as concluded previously.
While both groups of foundations are important and necessary we believe the former has a more direct and important role in the creation, evolution and sustainability of open source projects. Therefore, they will be the focus of the next research questions.  

\begin{framed}
\noindent 57 out of 101 foundations (56\%) which are involved in the creation of software product/s. 
Cross-checking this results with the ones obtained in RQ2, surveyed foundations mainly work to promote the development of software products their endorse either by supporting their communities or sponsoring the projects; among the others they are mainly devoted to promote the open and free software movement.
\end{framed}


\subsection{RQ4: Role in the Development Process}
\label{sec:results:rq3}

In this next step we go deeper in the analysis of some of the foundations in our dataset. 
In particular, we focus on the independent, international and transparent foundations aimed at supporting the development of a specific set of software products (i.e., those foundations marked as selected in the last three columns of Table~\ref{tab:foundations-rq1} plus including developing software as activity according to Table~\ref{tab:foundations-rq2}).

The first column in Table~\ref{tab:foundations-rq3} indicates the foundations considered, which are 27 in total.
We now complete the study of these foundations and look at, exactly, how these foundations aim to support their projects. 
Do they limit themselves to provide a legal and organizational framework for the software project? or do they go beyond this and try to dictate and monitor the governance and management of the software development process?

Our analysis covers five main attributes, namely: communication, code of conduct, the process to become a committer, the project governance and the existence of a technical board.
Table \ref{tab:foundations-rq3} summarizes the main results. 
Next we describe each dimension and report on the findings in more detail.

\begin{sidewaystable*}
\begin{threeparttable}
\caption{Analysis of the development process of the selected foundations. }
\label{tab:foundations-rq3}
\fontsize{7.0pt}{7.5pt}\selectfont
\begin{tabularx}{\textheight}{X@{\hspace{0.5em}}ccc@{\hspace{0.25em}}cc@{\hspace{0.5em}}c@{\hspace{0.5em}}cccccc@{\hspace{0.5em}}c@{\hspace{0.5em}}c} 
\toprule
\multirow{2}{*}{\textsc{Foundation}} &
\multicolumn{2}{c}{\textsc{Communication}} & \multirow{2}{*}{\textsc{CoC}} & &
\multicolumn{2}{c}{\textsc{Becoming a committer}} & &
\multicolumn{5}{c}{\textsc{Development Governance}} & &
\textsc{Technical} \\
\cmidrule{2-3}
\cmidrule{6-7}
\cmidrule{9-13}
&
\textsc{Main} & \textsc{Newcomers} & & &
\textsc{Who} & \textsc{How} & &
\textsc{Coord.} & \textsc{Impl.} & \textsc{Who} & \textsc{What} & \textsc{How} & & \textsc{Board} \\
\midrule
Apache Software Foundation \dotfill & 
    ML                     &   
    Y                      &   
    Y                      & & 
    Anyone                 &   
    Voucher\tnote{a}       & & 
    ML                     &   
    BZ/Jira                &   
    Members                &   
    Special\tnote{b}       &   
    Special\tnote{c}       & & 
    Y                      \\  
Cloud Foundry Foundation \dotfill & 
    ML, SL                 &   
    Y                      &   
    Y                      & & 
    Anyone                 &   
    Dojo                   & & 
    -                      &   
    -                      &   
    Project members        &   
    Special\tnote{d}       &   
    Consensus              & & 
    Y                      \\  
Django Software Foundation \dotfill & 
    ML                     &   
    Y                      &   
    Y                      & & 
    Anyone                 &   
    By request             & & 
    ML                     &   
    Trac                   &   
    Core devs.             &   
    Issues                 &   
    Special\tnote{e}       & & 
    Y                      \\  
Document Foundation \dotfill & 
    ML                     &   
    Y                      &   
    Y                      & & 
    -                      &   
    -                      & & 
    BZ                     &   
    Gerrit                 &   
    -                      &   
    Issues                 &   
    Voting                 & & 
    Y                      \\  
.NET Foundation \dotfill & 
    F                      &   
    Y                      &   
    Y                      & & 
    Anyone                 &   
    -                      & & 
    GH                     &   
    GH                     &   
    Anyone                 &   
    Issues, PR             &   
    Voting\tnote{f}        & & 
    Y                      \\  
Eclipse Foundation  \dotfill & 
    ML, F                  &   
    Y                      &   
    Y                      & & 
    Special\tnote{g}       &   
    Voting\tnote{h}        & & 
    BZ                     &   
    BZ                     &   
    Special\tnote{i}       &   
    Bugs                   &   
    Voting                 & & 
    Y                      \\  
Fintech Open Source Foundation  \dotfill & 
    -                      &   
    Y                      &   
    Y                      & & 
    Anyone                 &   
    By PR                  & & 
    GH                     &   
    GH                     &   
    Leader                 &   
    Issues, PR             &   
    BDFL                   & & 
    Y                      \\  
FreeBSD Foundation \dotfill & 
    F, ML, C               &   
    Y                      &   
    Y                      & & 
    -                      &   
    -                      & & 
    ML                     &   
    BZ                     &   
    -                      &   
    -                      &   
    -                      & & 
    -                      \\  
F\# Foundation  \dotfill & 
    IT                     &   
    Y                      &   
    Y                      & & 
    Members                &   
    Training               & & 
    GH                     &   
    GH                     &   
    Developers             &   
    RFCs                   &   
    Discussion             & & 
    Y                      \\  
Gentoo Foundation  \dotfill & 
    ML, F                  &   
    Y                      &   
    Y                      & & 
    Anyone                 &   
    Mentoring              & & 
    -                      &   
    -                      &   
    -                      &   
    -                      &   
    -                      & & 
    -                      \\  
GNOME Foundation  \dotfill & 
    ML, C                  &   
    Y                      &   
    Y                      & & 
    -                      &   
    -                      & & 
    BZ                     &   
    BZ                     &   
    Mentors                &   
    Issues                 &   
    BDFL                   & & 
    -                      \\  
Kuali Foundation  \dotfill & 
    -                      &   
    -                      &   
    -                      & & 
    -                      &   
    -                      & & 
    -                      &   
    -                      &   
    -                      &   
    -                      &   
    -                      & & 
    -                      \\  
Mozilla Foundation  \dotfill & 
    ML, F                  &   
    Y                      &   
    Y                      & & 
    Anyone                 &   
    Voucher\tnote{a}       & & 
    BZ                     &   
    BZ                     &   
    Special\tnote{j}       &   
    Issues                 &   
    BDFL                   & & 
    Y                      \\  
NetBSD Foundation  \dotfill & 
    ML                     &   
    Y                      &   
    -                      & & 
    -                      &   
    -                      & & 
    Custom\tnote{k}        &   
    Custom\tnote{k}        &   
    -                      &   
    -                      &   
    -                      & & 
    Y                      \\  
NLnet Labs Foundation  \dotfill & 
    -                      &   
    -                      &   
    -                      & & 
    -                      &   
    -                      & & 
    -                      &   
    -                      &   
    -                      &   
    -                      &   
    -                      & & 
    -                      \\  
Open Source Geospatial Foundation  \dotfill & 
    ML                     &   
    Y                      &   
    Y                      & & 
    -                      &   
    -                      & & 
    -                      &   
    -                      &   
    -                      &   
    -                      &   
    -                      & & 
    -                      \\  
OpenBSD Foundation  \dotfill & 
    ML                     &   
    -                      &   
    -                      & & 
    -                      &   
    -                      & & 
    Custom \tnote{k}       &   
    Custom \tnote{k}       &   
    -                      &   
    -                      &   
    -                      & & 
    -                      \\  
OpenSourceMatters  \dotfill & 
    ML                     &   
    Y                      &   
    Y                      & & 
    -                      &   
    -                      & & 
    ML, IT                 &   
    IT                     &   
    -                      &   
    Issues                 &   
    -                      & & 
    -                      \\  
OpenStack Foundation  \dotfill & 
    IRC, ML                &   
    Y                      &   
    Y                      & & 
    Special\tnote{l}       &   
    Paid                   & & 
    IT                     &   
    IT                     &   
    Core reviewers         &   
    Issues                 &   
    Voting\tnote{m}        & & 
    -                      \\  
OpenStreetMap Foundation  \dotfill & 
    IRC, ML                &   
    Y                      &   
    Y                      & & 
    -                      &   
    -                      & & 
    GH                     &   
    GH                     &   
    -                      &   
    Issues                 &   
    -                      & & 
    Y                      \\  
Parrot Foundation  \dotfill & 
    -                      &   
    -                      &   
    -                      & & 
    Special\tnote{g}       &   
    By request\tnote{n}    & & 
    ML, C                  &   
    GH                     &   
    -                      &   
    Issues                 &   
    -                      & & 
    Y                      \\  
Plone Foundation  \dotfill & 
    Forums                 &   
    Y                      &   
    Y                      & & 
    Anyone                 &   
    By PR                  & & 
    IT                     &   
    GH                     &   
    -                      &   
    Issues                 &   
    -                      & & 
    -                      \\  
Python Software Foundation  \dotfill & 
    ML                     &   
    Y                      &   
    Y                      & & 
    Anyone                 &   
    By PR                  & & 
    IT                     &   
    IT, ML                 &   
    Committers             &   
    Issues                 &   
    Discussion\tnote{o}    & & 
    Y                      \\  
Sahana Foundation  \dotfill & 
    Slack,ML               &   
    Y                      &   
    -                      & & 
    -                      &   
    -                      & & 
    -                      &   
    -                      &   
    -                      &   
    -                      &   
    -                      & & 
    -                      \\  
The Perl Foundation  \dotfill & 
    ML, C                  &   
    Y                      &   
    -                      & & 
    -                      &   
    -                      & & 
    ML                     &   
    -                      &   
    -                      &   
    -                      &   
    -                      & & 
    -                      \\  
Wikimedia Foundation  \dotfill & 
    ML, C, W               &   
    Y                      &   
    Y                      & & 
    Anyone                 &   
    Special\tnote{p}       & & 
    Phabricator            &   
    Gerrit                 &   
    Committers             &   
    Issues                 &   
    Discussion             & & 
    -                      \\  
X.Org Foundation LLC  \dotfill & 
    ML                     &   
    Y                      &   
    -                      & & 
    Special\tnote{g}       &   
    Sponsoring             & & 
    ML, Web                &   
    ML, BZ                 &   
    Reviewers              &   
    Patches                &   
    Discussion             & & 
    -                      \\  
\bottomrule
\end{tabularx}
\begin{tablenotes}
\item Legend: CoC = Code of Conduct, ML = Mailing-list, SL = Slack, F = Forums, C = Chat, W = Wiki, BZ = Bugzilla, IT = Ad-hoc Issue-Tracker, GH = GitHub, PR = Pull Request, BDFL = Benevolent Dictator For Life. Empty spaces mean that no information was gathered.
\item [a] The developer has to gather \emph{karma} (i.e., \emph{Apache Software Foundation}) or \emph{vouchers} (i.e., \emph{Mozilla Foundation}) from other developers.
\item [b] (a) code, (b) releases, (c) procedural.
\item [c] (a) code by absolute majority with vetoes, (b) releases require 3 binding votes in favour, (c) procedural by majority rule.
\item [d] Proposals are described as appeals that must have 25\% of support to be considered
\item [e] Four fifths majority (technical board has veto right).
\item [f] Original contributor or project leads can act as BDFL.
\item [g] Any developer with a clear track of contributors in the project.
\item [h] Committers vote (at least 3 votes are required) and have veto. The project committee approves/rejects the final decision.
\item [i] Committers and technical board.
\item [j] Bugzilla component owners, module owners, release drivers, ultimate decision-makers.
\item [k] By command line (GNATS system for NetBSD Foundation).
\item [l] Individual Foundation Members (vs. Individual Community members, who cannot commit)
\item [m] Core reviewers have +/-2 and W+1 rights, that is required for blocking or approving a patch
\item [n] Developer requests to be committer (or is nominated), architect approves and metacommitter applies.
\item [o] The final decision is overiden by Guido Van Rossum, author of the Python programming language.
\item [p] Committers can apply for it but \emph{+2} committers have to be selected after showing merits and passing internal hiring standards.
\end{tablenotes}
\end{threeparttable}
\end{sidewaystable*}

\vspace{\mysep}
\noindent\textbf{Communication}.
In OSS projects, communication plays a key role to disseminate the project and to help onboard new developers, also known as newcomers. 
For each foundation, we evaluate the main communication channels and the resources provided to help newcomers to participate in the daily life of the project (e.g., specific portals or indications to identify where help is required).

The most common communication channel is the mailing list, although the use of forums is also widely spread.
Mailing lists are generally used due to the ability to track the messages while forums are commonly used to record discussions or technical issues (and generally, forums are deployed together with mailing lists). The use of chat environments (e.g., IRC or Slack) has been detected scarcely. 

Regarding the resources provided for newcomers, the great majority of the foundations provide specific documentation to help their onboarding. 
This lowers the entry barrier to any developer willing to contribute to the software projects covered by these foundations, paving also the way to the future promotion of some of them to core project committers.

\vspace{\mysep}
\noindent\textbf{Code of Conduct}.
The community behind OSS projects is generally a mixture of people with a wide range of backgrounds, cultures, personalities and interests.
This rich mixture of people may increase the risk of offensive behaviors happening, thus creating bad working atmosphere and eventually leading to a drain of support and contributors in the project.
The concept of ``code of conduct'' was created as a means to protect community members from these negative behaviors.

The code of conduct of a software project defines a set of principles and values to expect from community member behavior.
It establishes both communication rules and enforcement mechanisms to allow anyone to feel safe and comfortable while contributing in the OSS project.

The results reveal that most surveyed foundations provided a code of conduct. 
We found the corresponding code of conduct file in 19 out of 27 (70\%) of the surveyed foundations.
The analysis of these files show that they aim at promoting a proper welcoming, safe, friendly and inclusive environment where members can easily understand how to behave.

\vspace{\mysep}
\noindent\textbf{Becoming committer}.
OSS projects capitalize on people contributing to their development (\cite{DBLP:journals/ais/Iivari09}). 
Hence, retaining and capturing developers is crucial for their success. 
This is especially true for committers who are developers that have write access to the codebase and therefore have the key role to write/review new code to be committed to the project. 

Our results reveal that becoming committer is in principle open to anyone interested in the project providing that has shown enough commitment and interest in it. 
The actual selection and evaluation process varies from foundation to foundation. 
Some of the decision-making mechanisms to select committers (see Table \ref{tab:foundations-rq3} for the full list) include: 
(1) use of vouchers or some mechanism to gather support from other developers (e.g., \emph{Apache Software Foundation} or \emph{Mozilla Foundation}), 
(2) a mentoring/training process (e.g., \emph{Gentoo Foundation} or \emph{F\# Foundation}) 
(3) or based on the number and success of previous pull-requests (e.g., \emph{Python Foundation} or \emph{Fintech Open Source Foundation}).

\vspace{\mysep}
\noindent\textbf{Governance}.
Effective and precise prioritization of the development tasks is also a crucial activity in any OSS project.
With this purpose, each foundation defines and applies its own set of governance rules to govern the projects under its umbrella. 
These governance rules describe how to contribute to the project and how decisions regarding the acceptance/rejection of such contributions are going to be made.
We study how people coordinate (see \textsc{Coord.} column), the tool/s they use to do so (see \textsc{Impl.} column) and how they govern the process (see \textsc{Who}, \textsc{What} and \textsc{How} columns).

As can be seen in the Table, the definition of governance models is maybe the most ignored issue in the analyzed foundations, even among the transparent ones.  
Some analyzed foundations are completely opaque (e.g., \emph{Kuali Foundation} or \emph{NLnet Labs Foundation}, where their projects are solely developed by their employees). For those that do provide some information about governance, information is typically not clearly available in one single place. 
Instead, governance information is scattered across several documents. 
According to our results, most of the foundations rely on issue-trackers to identify development tasks for their projects (see \textsc{What} column), however, it is often not clear how these tasks are prioritized, approved or refused (see \textsc{Who} and \textsc{How} columns). 

A remarkable exception is the \emph{Apache Software Foundation} that clearly describes the development process of the Apache projects part of the foundation. 
In Apache projects, committees have the decision-making authority regarding the content and direction of the Apache projects.
In turn, these committees are overseen by the foundation's board.
Decision making processes normally follow a lazy consensus approach: a few positive votes with no negative vote is enough for approval. 
Negative votes always include some explanation for the reason to reject and may include an alternative proposal to be discussed.

Another example is the \emph{Mozilla Foundation}, who describes how the development process is governed. 
The governance of Mozilla modules is a meritocracy, where developers need to demonstrate their abilities and find someone in the community who has adequate authority and will vouch for his/her competence.
Module development is governed by the module owners, while release management is decided by the release drivers.
The full governance process is overseen by super-reviewers, who review code for its effects on the overall state of the development branch and its adherence to Mozilla coding guidelines; and ultimate decision-makers, who are trusted members of the community who have the final say in the case of disputes.

We also found extreme cases regarding the presence or absence of indications from the foundation to its projects regarding how they should be governed. 
In some cases, the foundation explicitly states that it takes no part in the day-to-day activities of the project, like the statement done by the \emph{Open Source Geospatial Foundation}, which reads \emph{the foundation is not interested in controlling foundation projects}, thus clearly separating the foundation structure and the development process of its projects. 
However, even in those cases, the foundation may give some recommendations for the governance of its projects, like the rejection of the \emph{benevolent dictator for life} governance role or the enforcement of code management systems.
In other cases, the foundation bylaws explicitly refers to the governance model to be deployed in every project within the foundation, like the statement done by the \emph{.NET Foundation}, which reads \emph{Open source software projects within the Foundation will be subject to a governance process adopted by the Board of Directors}.
We checked that projects can later particularize the imposed governance model to their need, but we believe that these indications help to define a framework which unifies the governance within the foundation.

\vspace{\mysep}
\noindent\textbf{Technical Board}.
Some software projects may employ boards composed of recognized developers to drive the main technical decisions that may raise during the development process.
Becoming part of these boards is usually hard and requires demonstrating clear compromise and commitment to the project.
We study first whether these boards exist and then how people can be part of them.

Around half of the analyzed foundations make use of technical boards. 
Typically, these boards focus on technical aspects with an advisory role and are not used as a ``formal'' bridge between the project and the foundation internal organization. 
An exception is Apache and Eclipse, where the project committee (i.e., the Project Management Committees, PMCs) periodically report to the board about the status of the project.

\begin{framed}
\noindent Most of the foundations provide communication means and useful information for newcomers, but have limited implication and influence in the software project day-to-day governance and decision-making process. 
\end{framed}


\subsection{RQ5: Openness}
\label{sec:results:rq4}

As a final aspect of our study, we analyze whether the foundations themselves (and not the projects they support) adhere as well to an open philosophy. 
We focus on the same foundations studied in the previous research question.

To answer this question, we study their internal organization according to three main attributes, including the board, the membership and the meetings.
It turns out that the internal organization of the analyzed foundations follows always a similar pattern, with only a few exceptions. 
We report on this in the following. 

\vspace{\mysep}
\noindent\textbf{Board}. 
In an organization, the board of directors is usually a recognized group of people who is responsible for the management and oversight of the organization. 
The directors' powers, duties and responsibilities are generally specified in the bylaws of the foundation, where it is also common to specify the size of the board, who can be elected and how they are chosen or removed.
In our analysis, we study essential information to check the openness of the foundations in this regard, including its size, the term, who can be part of the board, how they make decisions and how they are elected/removed from their seats.

Table \ref{tab:foundations-rq4-board} summarizes this information. 
In general, the information about the board is openly available, where the board of directors has 1-year term and is elected by the members of the organization; besides, board actions are taken by majority with a majority quorum. 
Regarding the specificities, the analysis reveals that the decision-making mechanism applied for agreeing on actions in the board, electing and removing directors usually differs.
While a majority (or simple majority) is usually applied for taking action, a variety of different mechanisms are put in place for electing and removing directors. 
Also, the election of directors usually involves the members of the organization but their removal sometimes does not count on them (e.g., \emph{Django Foundation} or \emph{X.Org Foundation}), thus restricting the members' freedom.

\begin{sidewaystable*}
\begin{threeparttable}
\caption{Analysis of the board of the selected foundations. }
\label{tab:foundations-rq4-board}
\fontsize{7.0pt}{7.5pt}\selectfont
\begin{tabularx}{\textheight}{X@{\hspace{0.5em}}ccc@{\hspace{0.5em}}cc@{\hspace{0.5em}}c@{\hspace{0.5em}}ccc@{\hspace{0.5em}}c@{\hspace{0.5em}}ccc} 
\toprule
\multirow{2}{*}{\textsc{Foundation}} &
\multirow{2}{*}{\textsc{Size}} &
\multirow{2}{*}{\textsc{Term}} &
\multicolumn{2}{c}{\textsc{Actions}} & &
\multicolumn{3}{c}{\textsc{Election}} & &
\multicolumn{3}{c}{\textsc{Removal}}\\
\cmidrule{4-5}
\cmidrule{7-9}
\cmidrule{11-13}
 & & &
               \textsc{How} & \textsc{Quorum} & &
\textsc{Who} & \textsc{How} & \textsc{Quorum} & &
\textsc{Who} & \textsc{How} & \textsc{Quorum}  \\
\midrule
Apache Software Foundation  \dotfill & 
  $9    $                 & 
  1 year                  & 
  M                       & 
  M                       & & 
  Members                 & 
  M                       & 
  -                       & & 
  Members                 & 
  M                       & 
  -                       \\
Cloud Foundry Foundation  \dotfill & 
    $9$                     & 
    Until removal           & 
    SM                      & 
    P                       & & 
    Special\tnote{a}        & 
    SM                      & 
    P                       & & 
    Special\tnote{b}        & 
    SupM                    & 
    -                       \\
Django Software Foundation  \dotfill & 
  $3-15 $                 & 
  1 year                  & 
  M                       & 
  M                       & & 
  Members                 & 
  M                       & 
  -                       & & 
  Directors               & 
  M                       & 
  -                       \\
Document Foundation         \dotfill & 
    $7    $                 & 
    2 years                 & 
    SM                      & 
    Half                    & & 
    Trustees\tnote{c}       & 
    STV                     & 
  -                       & & 
  Trustees                & 
  STV                     & 
  -                       \\
.NET Foundation            \dotfill & 
    3                       & 
    1 year                  & 
    M                       & 
    M                       & & 
    Special\tnote{d}        & 
    Special\tnote{d}        & 
    -                       & & 
    Members                 & 
    Unanimously             & 
    -                       \\
Eclipse Foundation          \dotfill & 
    $> 1  $                 & 
    Vary\tnote{e}           & 
    SM\tnote{f}             & 
    SM                      & & 
    Vary\tnote{g}           & 
    Vary\tnote{h}           & 
  -                       & & 
  Directors               & 
  Vary\tnote{i}           & 
  SM\tnote{j}             \\
Fintech Open Source Foundation  \dotfill & 
    $21-29$                 & 
    Vary\tnote{t}           & 
    M                       & 
    M                       & & 
    Board\tnote{u}          & 
    M                       & 
  -                       & & 
  Board                   & 
  -                       & 
  -                       \\
FreeBSD Foundation          \dotfill & 
    $>=3$                   & 
    1 year                  & 
    M                       & 
    M                       & & 
    Board                   & 
    M                       & 
    M                       & & 
    -                       & 
    -                       & 
    -                       \\
F\# Foundation              \dotfill & 
    $3-19$                  & 
    1 year                  & 
    M                       & 
    M                       & & 
    Members                 & 
    -                       & 
    $1/4$                   & & 
    Board                   & 
    -                       & 
    -                       \\
Gentoo Foundation           \dotfill & 
    $5    $                 & 
    1 year                  & 
    SM                      & 
    M                       & & 
    Members                 & 
    Condorcet               & 
  -                       & & 
  Members                 & 
  M                       & 
  -                       \\
GNOME Foundation            \dotfill & 
    Not fixed               & 
    1 year                  & 
    M                       & 
    M                       & & 
    Members                 & 
    STV                     & 
  -                       & & 
  Board                   & 
  -                       & 
  -                       \\
Kuali                       \dotfill & 
    $5-7$                   & 
    3 years                 & 
    M                       & 
    M                       & & 
    Board                   & 
    M                       & 
    M                       & & 
    Board                   & 
    M                       & 
    M                       \\
Mozilla Foundation          \dotfill & 
    $5-15 $                 & 
    1 year                  & 
    M                       & 
    M                       & & 
    Board                   & 
    M                       & 
  M                       & & 
  Board                   & 
  M                       & 
  Full                    \\
NetBSD Foundation           \dotfill & 
    $3-9  $                 & 
    2 years                 & 
    SM                      & 
    M                       & & 
    Committee\tnote{k}      & 
    SM                      & 
  $25\%$                  & & 
  Active Members          & 
  $85\%$                  & 
  $25\%$                  \\
NLnet Labs Foundation           \dotfill & 
    $3-5  $                 & 
    3 years                 & 
    AbsM\tnote{l}           & 
    M                       & & 
    Board                   & 
    AbsM\tnote{l}           & 
    -                       & & 
    Board                   & 
    Unanimously             & 
    -                       \\
Open Source Geospatial Foundation  \dotfill & 
    $5-9  $                 & 
    1 year                  & 
    M                       & 
    M                       & & 
    Members                 & 
    M                       & 
  M                       & & 
  Vary\tnote{m}           & 
  Vary\tnote{n}           & 
  -                       \\
OpenBSD Foundation          \dotfill & 
    $>3   $                 & 
    1 year                  & 
    M                       & 
    -                       & & 
    Members                 & 
    M                       & 
  M                       & & 
  Members                 & 
  M                       & 
  M                       \\
OpenSourceMatters           \dotfill & 
    Special\tnote{o}        & 
    $1/2$ year              & 
    M                       & 
    M                       & & 
    Team leaders            & 
    M                       & 
  -                       & & 
  Members                 & 
  $2/3$                   & 
  -                       \\
OpenStack Foundation        \dotfill & 
    $<25$                   & 
    Until removal           & 
    M                       & 
    M                       & & 
    Members                 & 
    Plurality               & 
    M                       & & 
    Board                   & 
    -                       & 
    -                       \\
OpenStreetMap Foundation    \dotfill & 
    $2-8$                   & 
    Anually\tnote{p}        & 
    M                       & 
    M                       & & 
    Members                 & 
    Plurality               & 
    M                       & & 
    Board                   & 
    -                       & 
    -                       \\
Parrot Foundation           \dotfill & 
    $>2  $                  & 
    1 year                  & 
    M                       & 
    M                       & & 
    Members                 & 
    -                       & 
  -                       & & 
  Members                 & 
  M                       & 
  -                       \\
Plone Foundation            \dotfill & 
    $-    $                 & 
    1 year                  & 
    M                       & 
    M                       & & 
    Members                 & 
    M                       & 
  -                       & & 
  Members                 & 
  M                       & 
  -                       \\
Python Software Foundation  \dotfill & 
    $3-11 $                 & 
    Vary\tnote{q}           & 
    M                       & 
    M                       & & 
    Members                 & 
    Vote ordering\tnote{r}  & 
  -                       & & 
  Board                   & 
  $2/3$                   & 
  -                       \\
Sahana Foundation  \dotfill & 
    $5-15 $                 & 
    Vary\tnote{s}           & 
    M                       & 
    P                       & & 
    Members                 & 
    -                       & 
    -                       & & 
    Board/Members           & 
    $2/3$                   & 
    -                       \\
The Perl Foundation         \dotfill & 
    $> 1  $                 & 
    2 years                 & 
    M                       & 
    M                       & & 
    Board                   & 
    -                       & 
  -                       & & 
  Board                   & 
  M                       & 
  M                       \\
Wikimedia Foundation        \dotfill & 
    $>= 9 $                 & 
    3 years                 & 
    M                       & 
    M                       & & 
    Board                   & 
    M                       & 
    -                       & & 
    Board                   & 
    M                       & 
    -                       \\
X.Org Foundation LLC        \dotfill & 
    $< 8  $                 & 
    1-2 years               & 
    M                       & 
    M                       & & 
    Members                 & 
    Vote ordering\tnote{v}  & 
  -                       & & 
  Board                   & 
  $2/3$                   & 
  -                       \\
\bottomrule
\end{tabularx}
\begin{tablenotes}
\item Legend: M = Majority. SM = Simple Majority. SupM = Supermajority. AbsM = Absolute Majority. STV = Simple majority with single transferable vote + Meek Method. P = Present. Empty spaces mean that no information was gathered.
\item [a] Platinum members elect Platinum directors; the same applies for Gold.
\item [b] Board or at discretion by the member that appointed him/her.
\item [c] Trustees are developers who have verifiably contributed time and intellectual work.
\item [d] Members choose 2 directors by majority, any community participant choose 1 director by majority.
\item [e] (a) Strategic developer and consumer directors until their removal, (b) add-In directors and committer directors annually.
\item [f] (a) SM as default, (b) unanimous for license-related issues, (c) SupM for foundation-related issues, (d) unanimously when no meeting, (e) membership approval otherwise.
\item [g] Each developer group (strategic developer / strategic consumer / add-in provider / committer) appoints its director.
\item [h] Strategic developer and strategic consumer directors unanimously; add-in provider and committer directors by STV.
\item [i] (a) With cause by majority, (b) with no cause by two-thirds of the directors.
\item [j] Good Standing directors (or one-half when even number required).
\item [k] 10 voting volunteers from active members.
\item [l] Ties are resolved by a chairman casting vote.
\item [m] (a) Without cause: member entitled to vote, (b) with cause: board.
\item [n] (a) Without cause: two thirds, (b) with cause: majority of directors (except the one being removed).
\item [o] Number of departments.
\item [p] 1/3 of the board will retire at each general meeting (starting with the oldest ones)
\item [q] Cohort A directors have three (3) years beginning. Cohort B directors have two (2) years. Cohort C directors have for one (1) year.
\item [r] The most voted candidates cohort A directors, the next four most voted become cohort B directors, and the next three most voted become cohort C directors.
\item [s] 3 of them 1-year term; 2 of them 2-year term; 3 of them 3-year term
\item [t] Depends on director's type.
\item [u] Within their types.
\item [v] The 4 most voted members become directors for 2 years. The next most voted will fill the board up to 8 and will hold office for a term of 1 year.
\end{tablenotes}
\end{threeparttable}
\end{sidewaystable*}

Some of the analyzed foundations classify the organization memberships according to groups (i.e., \emph{Eclipse Foundation}, \emph{Cloud Foundry Foundation} and \emph{OpenStack Foundation}) which may have different selection processes or privileges assigned. 
The \emph{Eclipse Foundation} is maybe the most peculiar one.
It classifies the organization membership into four main groups, namely: (a) strategic developers, (b) strategic consumers, (c) add-in providers and (d) committers. 
Each group has the right to elect/remove a number of corresponding directors for the board.
The election for the first two groups is unanimously while for the last two groups is by single transferable vote. 
The removal, however, is usually by majority for the four groups.
Finally, the decision-making mechanism depends on the action being taken (see table footnotes), but simple majority is applied by default.
This board organization shows a customized structure for the roles of the community, which may promote a better management of the organization.

Other issues worth noting are how some foundations select a special group of people to manage the board and the definition of causes for removing board members.
Although they are called differently (e.g., \emph{trustees}, \emph{active members} or \emph{committee members}) they share a common objective and can be considered the \emph{elite} in the organization, as they have the power to modify the board.
On the other hand, some of the foundations clearly establish the procedures to follow to remove directors with/without a cause, thus proposing different decision-making mechanisms for each one.

\vspace{\mysep}
\noindent\textbf{Membership}. 
Most organizations admit people to become members, thus allowing them to participate in the decision-making processes like the election of the board or other affairs of the corporation.
We are interested in analyzing who can be part of the foundation membership as well as how they are elected/removed.

Table \ref{tab:foundations-rq4-members} shows the results of the analysis of the membership for the selected foundations.
Decisions on new members usually relies on the current members, who participate in the election and removal processes. 
The decision-making mechanism applied differs across the foundations. 
Although some of them apply the well-known majority procedure, most of them rely on specific procedures, for instance, nomination mechanisms (e.g., \emph{Open Source Geospacial Foundation} and \emph{Parrot Foundation}), or simple agreement (e.g., \emph{NetBSD Foundation} or \emph{Python Software Foundation} for basic member level).
It is also interesting to note six of the analyzed foundations rely on a paid-based mechanism to become a member of the foundations, thus restricting the membership only to those developers/entities able to afford the fee.

\begin{table*}[t]
\centering
\begin{threeparttable}
\caption{Analysis of the membership of the selected foundations. }
\label{tab:foundations-rq4-members}
\fontsize{6.8pt}{7.5pt}\selectfont
\begin{tabularx}{0.9\textwidth}{X@{\hspace{0.5em}}ccc@{\hspace{0.5em}}c@{\hspace{0.5em}}cccl} 
\toprule
\multirow{2}{*}{\textsc{Foundation}} &
\multicolumn{3}{c}{\textsc{Election}} & &
\multicolumn{3}{c}{\textsc{Removal}} &
\multirow{2}{*}{\textsc{Special}} \\
\cmidrule{2-4}
\cmidrule{6-8}
&
\textsf{Who} & \textsc{How} & \textsc{Quorum} & &
\textsf{Who} & \textsc{How} & \textsc{Quorum} & \\
\midrule
Apache Software Foundation\dotfill & 
  Members               & 
  M                     & 
  -                     & & 
  Members               & 
  $2/3$                 & 
  -                     & 
    -                     \\
Cloud Foundry Foundation  \dotfill & 
    Anyone                & 
    Paid                  & 
    -                     & & 
    Board                 & 
    At will               & 
    -                     & 
    -                     \\
Django Software Foundation\dotfill & 
    Board                 & 
    M                     & 
    -                     & & 
    Board                 & 
    M                     & 
    -                     & 
    -                     \\
Document Foundation       \dotfill & 
    Trustees\tnote{a}     & 
    STV                   & 
    -                     & & 
    Trustees\tnote{a}     & 
    STV                   & 
    -                     & 
    2 years term          \\
.NET Foundation           \dotfill & 
    Members               & 
    Unanimously           & 
    -                     & & 
    -                     & 
    -                     & 
    -                     & 
    -                     \\
Eclipse Foundation        \dotfill & 
    None                  & 
    Paid                  & 
    -                     & & 
    Board                 & 
    $2/3$                 & 
    -                     \\ 
Fintech Open Software Foundation  \dotfill & 
    -                     & 
    Paid                  & 
    -                     & & 
    Board                 & 
    SupM                  & 
    -                     & 
    -                     \\
FreeBSD Foundation        \dotfill & 
    -                     & 
    -                     & 
    -                     & & 
    -                     & 
    -                     & 
    -                     & 
    -                     \\
F\# Foundation            \dotfill & 
    -                     & 
    Paid                  & 
    -                     & & 
    Board                 & 
    Unanimously           & 
    -                     \\ 
Gentoo Foundation         \dotfill & 
    Trustees\tnote{a}     & 
    Special\tnote{b}      & 
    -                     & & 
    Trustees\tnote{a}     & 
    M                     & 
    -                     & 
    -                     \\
GNOME Foundation          \dotfill & 
    Board                 & 
    M                     & 
    -                     & & 
    Board                 & 
    M                     & 
    -                     & 
    2 years term          \\
Kuali Foundation          \dotfill & 
    Board                 & 
    Paid                  & 
    -                     & & 
    Board                 & 
    -                     & 
    -                     & 
    -                     \\
Mozilla Foundation        \dotfill & 
    None\tnote{c}         & 
    -                     & 
    -                     & & 
    -                     & 
    -                     & 
    -                     & 
    -                     \\
NetBSD Foundation         \dotfill & 
    None                  & 
    Agreement\tnote{d}    & 
    -                     & & 
    -                     & 
    -                     & 
    -                     & 
    -                     \\
NLnet Labs Foundation     \dotfill & 
    -                     & 
    -                     & 
    -                     & & 
    -                     & 
    -                     & 
    -                     & 
    -                     \\
Open Source Geospatial Foundation  \dotfill & 
    Members               & 
    Nomination\tnote{e}   & 
    -                     & & 
    Members               & 
    $2/3$                 & 
    -                     & 
    -                     \\
OpenBSD Foundation        \dotfill & 
    Board                 & 
    M                     & 
    -                     & & 
    -                     & 
    -                     & 
    -                     & 
    -                     \\
OpenSourceMatters         \dotfill & 
    Members               & 
    $2/3$                 & 
    -                     & & 
    Members               & 
    Unanimously           & 
    -                     & 
    -                     \\
Openstack Foundation      \dotfill & 
    Ex. Director          & 
    Special\tnote{f}      & 
    -                     & & 
    Board                 & 
    -                     & 
    -                     & 
    -                     \\
OpenStreetMap Foundation  \dotfill & 
    Board                 & 
    Paid                  & 
    -                     & & 
    Board                 & 
    -                     & 
    -                     & 
    -                     \\
Parrot Foundation         \dotfill & 
    Members               & 
    Special\tnote{i}      & 
    -                     & & 
    Committee             & 
    Time-based\tnote{j}   & 
    -                     & 
    -                     \\
Plone Foundation          \dotfill & 
    Board                 & 
    M                     & 
    -                     & & 
    Board                 & 
    $2/3$                 & 
    -                     & 
    1 year                \\
Python Software Foundation \dotfill & 
    Board                 & 
    Special\tnote{k}      & 
    Board                 & & 
    Members               & 
    $2/3$                 & 
    -                     & 
    -                     \\
Sahana Foundation \dotfill & 
    Members               & 
    Special\tnote{l}      & 
    -                     & & 
    Members               & 
    $2/3$                 & 
    -                     & 
    -                     \\
The Perl Foundation       \dotfill & 
    None\tnote{c}         & 
    -                     & 
    -                     & & 
    -                     & 
    -                     & 
    -                     & 
    -                     \\
Wikimedia Foundation      \dotfill & 
    None\tnote{c}         & 
    -                     & 
    -                     & & 
    -                     & 
    -                     & 
    -                     & 
    -                     \\
X.Org Foundation LLC      \dotfill & 
    None                  & 
    Agreement\tnote{d}    & 
    -                     & & 
    Board                 & 
    $3/4$                 & 
    -                     & 
    -                     \\
\bottomrule
\end{tabularx}
\begin{tablenotes}
\item Legend: M = Majority. SM = Simple Majority. SupM = Super Majority. STV = Simple majority with single transferable vote + Meek Method.
\item Empty spaces mean that no information was gathered.
\item [a] Trustees are developers who have verifiably contributed time and intellectual work.
\item [b] Any active developer can become member unles there is an absolute majority of trustees that opposes it.
\item [c] There shall no be members.
\item [d] Signing a developer agreement.
\item [e] Nomination by other member and affirmative vote of the members.
\item [f] Request to Secretary, admited by the Executive Director
\item [i] Two nominations and affirmative vote of the board of directors.
\item [j] The Membership Committee may terminate any membership, after a period of 12 months without contribution or voting activity from the member.
\item [k] Depends of the member type, for basic level it's immediate. There are five levels, namely: (1) basic members, (2) supporting members, (3) sponsor members, (4) managing members, (5) contributing members and fellows.
\item[l] By nomination of members: (1) 20\% of Members submit a vote; (2) 75\% of those vote to admit the nominee; and (3) after receipt by the Secretary of a membership application completed by each such proposed Member within thirty (30) days following the vote.
\end{tablenotes}
\end{threeparttable}
\end{table*}



\vspace{\mysep}
\noindent\textbf{Meetings}. 
Most foundations hold an annual meeting, where members and the board can manage the business and affairs of the corporation.
These meetings are also used to renew the board and elect other possible committees of the foundation.
We study how frequent these meetings are and who can take part in them (Table \ref{tab:foundations-rq4-meetings}).
In the vast majority of foundations, any member can participate in the meeting and meeting decisions are based on a majority procedure with a quorum. 
However, most bylaws do not specify whether decisions made in these meetings can involve aspects that directly affect the development processes of their OSS projects. 
Only the \emph{Apache Software Foundation} defines the purpose of the meetings, which includes a revision of the status of every project, the election of developers for the committees or the approval of releases, among others. 

\begin{table*}[th]
\centering
\begin{threeparttable}
\caption{Analysis of the meetings hold by the selected foundations. }
\label{tab:foundations-rq4-meetings}
\fontsize{6.8pt}{7.5pt}\selectfont
\begin{tabularx}{0.75\textwidth}{X@{\hspace{0.5em}}ccccc} 
\toprule
\multirow{2}{*}{\textsc{Foundation}} &
\multirow{2}{*}{\textsc{Annual}} &
\multirow{2}{*}{\textsc{Special}} &
\multicolumn{3}{c}{\textsc{Actions}} \\
\cmidrule{4-6}
& & & 
\textsc{Who} & \textsc{How} & \textsc{Quorum} \\
\midrule
Apache Software Foundation    \dotfill & 
  Y                         & 
  Y                         & 
  Members                   & 
  M                         & 
  $1/3$                     \\
Cloud Foundry Foundation      \dotfill & 
    Y                         & 
    Y                         & 
    Board+Members\tnote{a}    & 
    Special\tnote{b}          & 
    P                         \\
Django Software Foundation    \dotfill & 
    Y                         & 
    Y                         & 
    Members                   & 
    M                         & 
    M                         \\
Document Foundation           \dotfill & 
    -                         & 
    -                         & 
    -                         & 
    -                         & 
    -                         \\
.NET Foundation               \dotfill & 
    Y                         & 
    Y                         & 
    Members                   & 
    M                         & 
    M                         \\
Eclipse Foundation            \dotfill & 
    Y                         & 
    Y                         & 
    Members                   & 
    SM\tnote{c}               & 
    SM\tnote{c}               \\
Fintech Open Source Foundation  \dotfill & 
    Y                         & 
    Y                         & 
    Platinum Members\tnote{h} & 
    M                         & 
    $51\%$                    \\
FreeBSD Foundation            \dotfill & 
    Y                         & 
    Y                         & 
    Board                     & 
    M                         & 
    M                         \\
F\# Foundation                \dotfill & 
    Y                         & 
    Y                         & 
    Members                   & 
    SM                        & 
    SM                        \\
Gentoo Foundation             \dotfill & 
    Y                         & 
    Y                         & 
    Members                   & 
    M                         & 
    $1/3$                     \\
GNOME Foundation              \dotfill & 
    Y                         & 
    Y                         & 
    Members                   & 
    M\tnote{d}                & 
    -                         \\
Kuali Foundation              \dotfill & 
    Y                         & 
    Y                         & 
    Board                     & 
    M                         & 
    M                         \\
Mozilla Foundation            \dotfill & 
    Y                         & 
    Y                         & 
    Board                     & 
    M                         & 
    M                         \\
NetBSD Foundation             \dotfill & 
    Y                         & 
    Y                         & 
    Active Members            & 
    M                         & 
    $1/4$                     \\
NLnet Labs Foundation             \dotfill & 
    Y                         & 
    Y                         & 
    Board                     & 
    AbsM                      & 
    M                         \\
Open Source Geospatial Foundation  \dotfill &  
    Y                         & 
    Y                         & 
    Members                   & 
    M                         & 
    M                         \\
OpenBSD Foundation            \dotfill & 
    Y                         & 
    Y                         & 
    Members                   & 
    M                         & 
    M                         \\
OpenSourceMatters             \dotfill & 
    Y                         & 
    -                         & 
    Members                   & 
    M                         & 
    M                         \\
OpenStack Foundation          \dotfill & 
    Y                         & 
    Y                         & 
    Members                   & 
    M                         & 
    10\%                      \\
OpenStreetMap Foundation      \dotfill & 
    Y                         & 
    Y                         & 
    Members+Board             & 
    Special\tnote{e}          & 
    Special\tnote{f}          \\
Parrot Foundation             \dotfill & 
    Y                         & 
    Y                         & 
    Members                   & 
    -                         & 
    $1/10$                    \\
Plone Foundation              \dotfill & 
    Y                         & 
    Y                         & 
    Members                   & 
    M                         & 
    M                         \\
Python Software Foundation    \dotfill & 
    Y                         & 
    Y                         & 
    Members\tnote{g}          & 
    M                         & 
    $1/3$                     \\
Sahana Foundation             \dotfill & 
    Y                         & 
    Y                         & 
    Members+Board             & 
    M                         & 
    P                         \\
The Perl Foundation           \dotfill & 
    Y                         & 
    Y                         & 
    Board                     & 
    M                         & 
    M                         \\
Wikimedia Foundation          \dotfill & 
    Y                         & 
    Y                         & 
    Board                     & 
    M                         & 
    M                         \\
X.Org Foundation LLC          \dotfill & 
    Y                         & 
    Y                         & 
    Members                   & 
    M                         & 
    $1/4$                     \\
\bottomrule
\end{tabularx}
\begin{tablenotes}
\item Legend: M = Majority. SM = Simple Majority. AbsM = Absolute Majority. P = Present
\item Empty spaces mean that no information was gathered.
\item [a] Members only upon Board's approval.
\item [b] Simple majority, Supermajority or Special Supermajority are applied depending on the topic.
\item [c] Each member has 1 vote (except committers of the same organization, who will have only 1). 
\item [d] The vote is usually cast by voice (or by ballot when required by the chairman). Fractional votes are not allowed
\item [e] Show of hands unless a poll is duly demanded.
\item [f] Either 15 members or 1/10 of the membership.
\item [g] Basic members are not included.
\item [h] There are four member levels, according to the fee paid, namely: Platinum, Gold, Silver, At-Large.
\end{tablenotes}
\end{threeparttable}
\end{table*}

\begin{framed}
\noindent The analyzed foundations show a high level of openness with most decision procedures based on member voting and democratic practices. Still, for a non-negligible number of foundations, participation in important decisions is restricted to paying members. 
\end{framed}


\subsection{Additional Discussion Points}
\label{sec:results:discussion}

Beyond the main findings reported so far, we would like to highlight additional contributions derived from the results.

\vspace{\mysep}
\noindent \textbf{Utility of umbrella foundations for new projects}.  
As we have seen, some foundations serve as umbrella for other smaller foundations or sets of related projects.
For instance, the \emph{Linux Foundation} clearly states that it \emph{provides unparalleled support for open source communities through financial and intellectual resources, infrastructure, services, events, and training}.
From the information we retrieved in our survey, this kind of foundations is especially useful for relatively young OSS projects as they provide the required scaffolding for their growth and liberate them from organizational issues while freeing them also from the burden of setting up their own foundation. 
Also, they create an ecosystem where projects and developers can easily collaborate together, which may help them to boost their development.

\vspace{\mysep}
\noindent \textbf{Weak alignment between the foundation and the project's concrete development practices}.  
Our survey has revealed that, while the organization of the foundation itself is well defined, this organization does not generally extend to the software projects that depend on it. 
As commented before, as projects grow, there is a need to put in place adequate structures to manage the community around the projects and optimize its contributions.
We expected foundations could take care of this issue, however, it turns out it does not seem to be the case in general.
We believe that a tighter integration between the foundations and the projects could help projects benefit from the organizational knowledge available at the foundation to decide the best governance and contribution model for the projects. 
This does not necessarily mean that projects should strictly follow the internal organization of the foundation itself but that at least the projects could use some of the know-how available there.  


\vspace{\mysep}
\noindent \textbf{Lack of precise documentation}.  
One of the main issues we found was the lack of precise documentation about how the foundation works. 
This may scare away some potential contributors that want to first clearly understand how their effort (e.g., in terms of time invested in coding patches for reported bugs) will be evaluated. 
Conversely, sometimes the information was abundant but unclear and scattered (e.g., the \emph{Wikimedia Foundation}), thus causing over-documentation issues and confusion, especially since it is likely that at some point this redundant documentation becomes partially outdated and even contradictory.
To maximize the efficiency and impact of foundations, we believe a clear and concise information about all foundation aspects is a must. 

\vspace{\mysep}
\noindent \textbf{No historical data publicly available}. 
Most of the foundations analyzed in our survey (with very few exceptions, like the \emph{Apache Software Foundation}) do not provide easy means to access the assets tracking the foundation activity (e.g., minutes of meetings, past composition of committees, historical list of projects supported, etc.).
This makes it really difficult to evaluate the real impact the organization has on open source since we can only have the current snapshot. 
No longitudinal studies can be done at this point. 
We encourage foundations to make the effort to open more of its data for further analysis. 

\vspace{\mysep}
\noindent \textbf{Towards more democratic development models}. 
We have observed that the use of voting mechanisms and promoting discussions is becoming increasingly common in the decision-making processes involved in software foundations, in particular, in their development processes.
Although we still find a lack of precise descriptions about governance rules, this observation may reveal the use of development models which are more democratic and go beyond the conventional Benevolent Dictator For Life (BDFL) model. 
For instance, the \emph{Python Foundation}, which relied on discussions to make decisions but Guido Van Rossum (the language creator) always had the last word (i.e., an implicit BDFL model); has been discussing new governance models for the project\footnote{https://www.python.org/dev/peps/pep-8000/} where community members have a key role when making decisions. 


\subsection{Replicability and Website}
\label{sec:results:replicability}

To facilitate the replication and validation of our survey, we have prepared a complete replication repository\footnote{http://hdl.handle.net/20.500.12004/1/J/ARXIV/2020/001} for interested researchers. 
The repository includes the list and results of the datasets used in the survey.

\begin{figure}[t]
\centering
\includegraphics[width=\columnwidth]{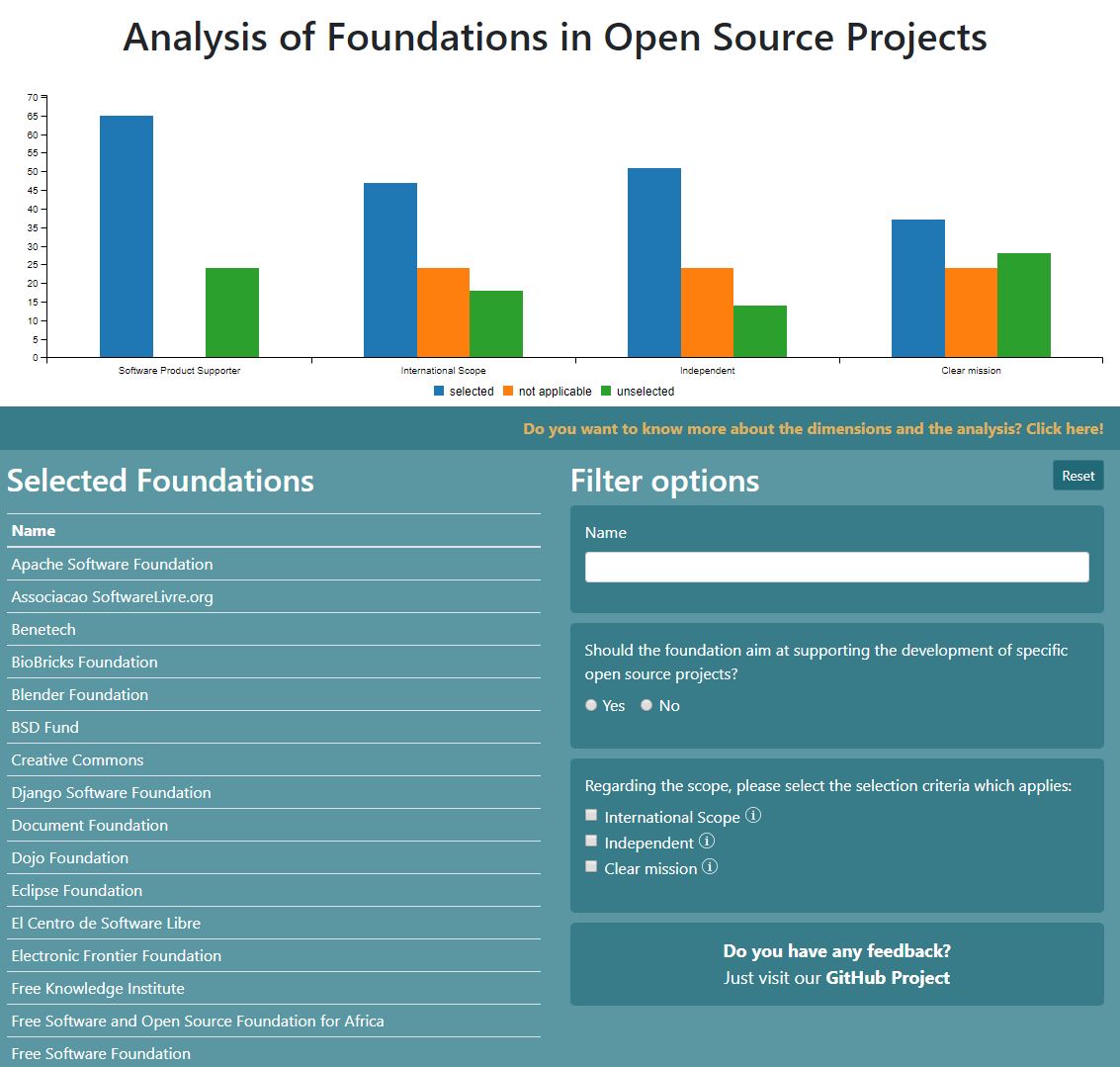}
\caption{Website developed.}
\label{fig:screenshot}
\end{figure}

Additionally we have developed a website\footnote{https://som-research.github.io/OSSFoundations} (available from the same repository) which offers an easy way to query and visualize the results of this survey. 
Figure~\ref{fig:screenshot} shows a screenshot of the website.
The results are presented in both a bar graph and a list, which can be queried according to the foundation name or previous dimensions. 
Each entry in the list shows the name, status form and dimension values of the selected foundation. 
We believe that the website may help any developer to easily query/filter the dataset and discover foundations that best suit their projects.
The website also serves as a living solution as any developer can contribute and recommend new foundations to consider in the dataset, as it has happened already.

\section{Threats to Validity}
\label{sec:threats}
Our work is subjected to a number of threats to validity, namely: 
(1) internal validity, which is related to the inferences we made; 
and (2) external validity, which discusses the generalization of our findings.
 
Regarding the internal validity, the dataset construction process starts from the set of foundations identified by \texttt{flossfoundations.org}.
To minimize the risk of missing foundations not listed there, we explored other online sources, our own previous knowledge and conversations with participants in open source conferences (like SustainOSS\footnote{https://sustainoss.org/}) and feedback we got after releasing the first version of the website accompanying this work to extend the initial list. 

Another threat to validity is our subjectivity in selecting and classifying the foundations of our dataset. 
A wrong perception or misunderstanding on our side may have resulted in a misclasification of foundations regarding the reported research questions.
To minimize this threat we applied a coding scheme as follows.
Data collection, selection and analysis were performed by one coder, who was the first author of this paper.
To ensure the quality of the process, a second coder, who was the second author of this paper, validated the input/output of each research question by randomly selecting a sample of size 30\% of the input of each one (e.g., a sample of 34 foundations for the first research question) and performing the selection and analysis himself.
The intercoder agreement reached for all research questions was higher than 80\%. 
Moreover, all disagreement cases were discussed between the coders to reach consensus and, when needed, other potentially affected foundations were reanalyzed in light of the refined analysis perspective. 

As for the external validity, note that our study is based on the sample of software foundations we constructed and therefore our results should not be generalized to other kinds of foundations linked to (non-software) collaborative projects, like book writing efforts or similar initiatives also frequent in GitHub or other code hosting platforms. 

\section{Related Work}
\label{sec:relatedWork}
In order to understand the context of our research work, the following subsections present previous research works on a number of relevant related topics: OSS development, OSS organizations \& foundations and governance in OSS. We discuss the relationship of our work with them at the end of the section.

\subsection{General Research on OSS development}
Research on OSS development has been widely addressed in the scientific literature especially given the growing number of organizations adopting open source practices (\cite{Spinellis2012}). ~\cite{Crowston2012} review the empirical research on Free/Libre and Open-Source Software (FLOSS) development and assess the state of the literature. 
Likewise, \cite{CosentinoIC17} perform a systematic mapping study of all research works studying software development aspects in the context of the GitHub code hosting platform. 
As an example, there are a number of works studying development models (\cite{Pinzger2014}), diversity of developer teams (\cite{VasilescuPRBSDF15}), community structures (\cite{Bird2008}), documentation procedures (\cite{Aggarwal2014}) or collaboration practices (\cite{Dabbish2012}), among many others. 

\subsection{Research on OSS Organizations \& Foundations}
Previous works on the topic of OSS organizations have focused on the study of the main elements characterizing such organizations and how they often end up evolving towards a foundation-like model.
For instance, a comparative study between the elements of OSS communities and the general elements of an organization (i.e., people, goals and organizational hierarchy) has been presented by \cite{DBLP:conf/oss/Eckert18}, thus helping to align better OSS and general-purpose organizations. 
Also, the work by \cite{DBLP:journals/jss/EckertSM19} analyzes how different organizational approaches (ranging from those ones operating beneath an umbrella organization to entirely inpendent ones) can be applied in OSS.
The work discusses the cases of four OSS organizational forms and gives some useful recommendations for other projects, as no one-size-fits-all solution exists. 

The evolution of the organization structure of OSS projects towards the creation of a foundation has been studied by works such as \cite{DBLP:conf/oss/RiehleB12}, which evaluates six foundations to build a comparative model that should help project leaders understanding the differences between existing foundations. 
The work by \cite{DBLP:conf/oss/LindmanH17} performs a qualitative analysis of how foundations support OSS projects, while the works of \cite{mahony} and \cite{DBLP:journals/computer/Riehle10} discuss the role of software foundations in OSS from a more economic/business perspective.
Some works have carefully analyzed specific projects and the foundation supporting them. For instance, the work by \cite{DBLP:journals/sopr/German03} studies the Gnome project (and the so-called foundation), its development methods and practices, and its organizational structure. 
The work by \cite{DBLP:journals/iepol/SadowskiSD08} follows a similar approach for the Debian project/foundation. 
These kind of works are especially useful to deeply understand how particular projects evolve and potential extrapolations to other projects.

\subsection{Governance in OSS}
Of special interest for our work is also the study of the (self-)organization of developer communities which eventually conforms some kind of governance model, for instance, by analyzing the social ties (\cite{Allaho2013}) and interactions (\cite{DBLP:journals/ese/GharehyaziePVF15}), or the blossoming of spontaneous collaboration networks (\cite{Bird2008,Bird11,Crowston2007}) as the project grows.
The definition of central concepts in governance, such as general characteristics of the collaboration or common life cycle stages in OSS communities, has been studied by \cite{lattermann2005}. 
The formalization of such collaboration is typically done by means of defining the governance model of the project (\cite{mahony2007,OMahony2007,Markus2007}) or even a governance index (\cite{Laffan2012}). 
Nevertheless, the explicit definition of governance models is still mostly missing in OSS projects, as analyzed by \cite{IzquierdoC15}. 

Additionally, a number of works have studied different coordination mechanisms that may be put in place to organize the work (\cite{Herbsleb1999,Kraut1995,CrowstonWLEH05}) or the mechanisms to protect their collaborative work (\cite{mahony2003}). 
The evolution over time of governance models has also been covered by works such as \cite{Panichella14}, \cite{Joblin17} and~\cite{DBLP:journals/tse/CapraFM08}. 

\subsection{Discussion}
Our exploration of the related literature shows that few works have actually surveyed OSS foundations considering the dimensions we covered. 
Among the few exceptions, we would like to mention the works of \cite{mahony} and \cite{DBLP:journals/computer/Riehle10}, the studies of \cite{DBLP:conf/oss/LindmanH17} and \cite{DBLP:conf/oss/RiehleB12}. 
While these studies (especially the last two) are useful to complement, from a more qualitative point of view, our RQ4 and RQ5 research questions, they restrict their analysis to a much smaller set of foundations and dimensions. 
As far as we know, ours is the first broad survey of software foundations and the role they play in open source. 

Some of the dimensions studied in this paper have been individually addressed by other works but at the project level. 
For instance, \cite{Tourani2017} study the importance of codes of conduct in OSS, where instructions on how to be part of the project community or to become committer are usually included.
\cite{GuzziBLPD13} study the role of communication channels in the development process of OSS, in particular, the use of mailing lists.
\cite{Prattico2012} carefully analyzes the bylaws of 6 OSS foundation and studies who holds the power in OSS software. 
We believe interesting to observe whether the study of the dimensions at the foundation level reports similar results.\looseness-1

\section{Conclusion}
\label{sec:conclusion}
We have performed a survey of 101 software foundations to better understand the role they play in open source development either at the legal, financial, marketing or development level. 
Our results show that most foundations focus on providing legal support and leading evangelizing actions while only a few few aim for a more complete support ``package'' including as well other aspects such as governance guidelines, code of conduct policies or development advice for the projects under their umbrella. 

As such, while we do believe that joining a foundation could be beneficial for open source projects to get expert support on topics like the ones pointed above where it may be difficult to find enough internal knowledge, this does not eliminate the need to setup clear and transparent contribution, governance and development processes at the project level. 
In fact, we believe projects should reach a certain level of growth and maturity before joining a foundation. 
This way, they will also be able to know better what kind of foundation is the best fit for them. 
In this sense, the website accompanying this work could be helpful to choose the right foundation for a project as it allows to display and filter the foundations according to a number of dimensions.  

As further work, we would like to compare the role foundations play in open source with the role they have in other kinds of non-governmental organizations (and non-profit organizations in general). 
Since foundations in open source are still rather young (compared to foundations in other domains), we hope that by learning from more established foundations in other fields, we can bring back some useful recommendations that maximize the impact software foundations may have in open source development, and especially in the long-term sustainability of open source. 
Furthermore, a follow-up qualitative study, comprising interviews with actual users and contributors of open source projects will also help to explore their views on the needs and expectations they have from software foundations and align better software projects and software foundations.



\IEEEtriggeratref{37}

\bibliographystyle{IEEEtran}
\bibliography{IEEEabrv,2020-Foundations}

\end{document}